\documentclass[preprint,nocopyrightspace]{sigplanconf}

\usepackage{amsmath}
\usepackage{amssymb}
\usepackage{mathpartir}
\usepackage{amsfonts}
\usepackage{stmaryrd}
\usepackage{MnSymbol}
\usepackage{code}
\usepackage{tikz}
\usetikzlibrary{matrix,arrows,decorations.pathmorphing}

\usepackage{listings}
\usepackage{xspace}
\usepackage{xcolor}
\usepackage{colortbl}
\usepackage{graphicx}
\usepackage{verbatim}
\usepackage{textcomp}
\usepackage[colorlinks=true]{hyperref}
\usepackage{mathrsfs}
\usepackage{syntax}
\newenvironment{round}
  {\begin{tikzpicture}[baseline={([yshift={-\ht\strutbox}]current bounding box.west)},
                    outer sep=0pt,inner sep=0pt]}
  {\draw [rounded corners=.5em] (table.north west) rectangle (table.south east);
   \end{tikzpicture}}

\newtheorem{theorem}{Theorem}
\newtheorem{corollary}{Corollary}
\newtheorem{proposition}[theorem]{Proposition}

\newtheorem{definition}[theorem]{Definition}
\newtheorem{remark}[theorem]{Remark}

\definecolor{shade}{RGB}{223,223,223}

\newcommand\shaderow{\noalign{\vskip-0.4pt}\rowcolor{shade}}

\newcommand{\trsim}{\lesssim}
\newcommand{\trbisim}{\eqsim}

\newcommand\refsec[1]{\hyperref[#1]{section \ref*{#1}}}

\lstdefinelanguage{links}{
  morekeywords={var,for},
  otherkeywords={<-,<--,=},
  morekeywords=[2]{table,database,with,from},
  classoffset=0,
  morestring=[b]"
}

\newcommand\ocamlc[1]{\lstinline[language={[Objective]Caml},basicstyle=\ttfamily\normalsize]{#1}}

\newcommand{\key}[1]{\textsf{#1}}
\newcommand{\fld}[1]{\key{#1}}

\newcommand{\str}[1]{\textsf{``#1''}}

\newcommand{\ampersand}{\mathbin{\textsf{\&}}}

\newcommand{\blist}{\begin{array}[t]{@{}l@{}}}
\newcommand{\elist}{\end{array}}

\newcommand{\EC}{\mathcal{E}}
\newcommand{\QC}{\mathcal{Q}}

\newcommand{\sem}[1]{\llbracket{#1}\rrbracket}

\newcommand{\recName}{\mathsf{rec}}
\newcommand{\rec}[3]{\recName~{#1}~{#2}.{#3}}

\newcommand{\Int}{\mathit{Int}}
\newcommand{\Bool}{\mathit{Bool}}
\newcommand{\String}{\mathit{String}}

\newcommand{\recordl}{\langle}
\newcommand{\recordr}{\rangle}
\newcommand{\record}[1]{\recordl {#1} \recordr}
\newcommand{\tuple}{\record}
\newcommand{\collection}[1]{\mathopen{\texttt{[}} {#1} \mathclose{\texttt{]}}}
\newcommand{\singleton}[1]{\ret{#1}} 
\newcommand{\concat}{\mathbin{+\!\!+}}

\newcommand{\Database}{\mathsf{database}}
\newcommand{\Query}{\mathsf{query}}

\newcommand{\Lift}{\mathsf{lift}}

\newcommand{\retl}{\texttt{[}}
\newcommand{\retr}{\texttt{]}}
\newcommand{\ret}[1]{\retl{#1}\retr}

\renewcommand{\plus}{\concat}
\newcommand{\zero}{\texttt{[]}}

\newcommand{\Let}{\mathsf{let}}
\newcommand{\In}{\mathsf{in}}

\newcommand{\Table}{\mathsf{table}}

\newcommand{\For}{\mathsf{for}}
\newcommand{\If}{\mathsf{if}}

\newcommand{\True}{\mathsf{true}}
\newcommand{\False}{\mathsf{false}}

\newcommand{\Quote}[1]{\mathopen{\texttt{<@}~} {#1} \mathclose{~\texttt{@>}}}
\newcommand{\AntiQuote}[1]{\texttt{(\%}{#1}\texttt{)}}

\newcommand{\eff}{\mathbin{!}}

\newcommand{\TO}[1]{\to^{#1}}

\newcommand{\db}{\mathbf{db}}
\newcommand{\pl}{\mathbf{pl}}
\newcommand{\any}{\mathbf{any}}

\newcommand{\envAt}[2]{\sem{#1}_{#2}}

\newcommand{\lookup}[3]{{#1} \mathbin{@_{#2}} #3}

\newcommand{\op}{\mathit{op}}
\newcommand{\opdelta}{\delta}

\newcommand{\tQuote}[1]{\mathsf{Expr}\mathopen{\texttt{<}}\,{#1}\,\mathclose{\texttt{>}}}
\newcommand{\tColl}[1]{\texttt{[}#1\texttt{]}}

\newcommand{\ba}{\begin{array}}
\newcommand{\ea}{\end{array}}

\newenvironment{mysyntax}{\[\ba{@{}l@{~~}l@{~}c@{~}l@{}}}{\ea\]}

\newenvironment{equations}{\[\ba{r@{~~}c@{~~}l@{\quad}l}}{\ea\]}

\newcommand{\cat}[1]{\text{({#1})}}

\newcommand{\TLINQ}{T-LINQ\xspace}

\newcommand{\Quot}{\mathsf{Quot}}
\newcommand{\QuotVar}{\mathsf{Quot}'}
\newcommand{\Eff}{\mathsf{Eff}}

\newcommand{\EffSub}{\mathsf{Eff_{\leq}}}

\newcommand{\dup}[1]{\langlebar{#1}\ranglebar}

\newcommand{\dupEff}[2]{\langlebar{#1}\ranglebar_{#2}}
\newcommand{\dupX}[1]{\dupEff{#1}{X}}
\newcommand{\dupY}[1]{\dupEff{#1}{Y}}
\newcommand{\duppl}[1]{\dupEff{#1}{\pl}}
\newcommand{\dupdb}[1]{\dupEff{#1}{\db}}

\newcommand{\trans}[1]{\sem{#1}}
\newcommand{\untrans}[1]{\llparenthesis{#1}\rrparenthesis}

\newcommand{\untransX}[1]{\untrans{#1}_X}
\newcommand{\untranspl}[1]{\untrans{#1}_{\pl}}
\newcommand{\untransdb}[1]{\untrans{#1}_{\db}}

\newcommand{\evalto}{\longrightarrow}

\newcommand{\rewriteto}{\leadsto}
\newcommand{\adhocto}{\hookrightarrow}

\newcommand{\normEff}[1]{|#1|}
\newcommand{\normQuot}[1]{\|#1\|}

\newcommand{\reify}{\mathord{\downarrow}}

\newcommand{\lift}[1]{\Lift~{#1}}

\newcommand{\fold}{\mathsf{fold}}

\newcommand{\transX}[1]{\trans{#1}_X}
\newcommand{\transpl}[1]{\trans{#1}_\pl}
\newcommand{\transdb}[1]{\trans{#1}_{\db{\vphantom{\any}}}}

\newcommand{\transXr}[1]{\transX{#1}^\rho}
\newcommand{\transplr}[1]{\transpl{#1}^\rho}
\newcommand{\transdbr}[1]{\transdb{#1}^\rho}

\newcommand{\error}{\mathit{error}}

\newcommand{\evaltoeff}[1]{\mathbin{\evalto}} 
\newcommand{\evaltoX}{\evaltoeff{X}}

\newcommand{\evaltopl}{\evaltoeff{\pl}}

\newenvironment{nop}{}{}
\newenvironment{smathpar}{
\begin{nop}\small\begin{mathpar}}{
\end{mathpar}\end{nop}\ignorespacesafterend}

\newcommand{\ltoo}[1]{\stackrel{#1}{\longrightarrow}}
\newcommand{\lToo}[1]{\stackrel{#1}{\Longrightarrow}}
\newcommand{\Too}{\Longrightarrow}
\newcommand{\too}{\longrightarrow}

\newcommand{\unit}{{\record{}}}

\title{Effective Quotation}
\subtitle{Relating approaches to language-integrated query}

\authorinfo{James Cheney\and Sam Lindley}{The University of
  Edinburgh}{jcheney@inf.ed.ac.uk, Sam.Lindley@ed.ac.uk}
\authorinfo{Gabriel Radanne}{ENS Cachan}{gabriel.radanne@zoho.com}
\authorinfo{Philip Wadler}{The University of Edinburgh}{wadler@inf.ed.ac.uk}
\begin{document}


\maketitle

\begin{abstract}
Language-integrated query techniques have been explored in a number of
different language designs. We consider two different, type-safe
approaches employed by Links and F\#. Both approaches provide rich
dynamic query generation capabilities, and thus amount to a form of
heterogeneous staged computation, but to date there has been no formal
investigation of their relative expressiveness. We present two core
calculi Eff and Quot, respectively capturing the essential aspects of
language-integrated querying using effects in Links and quotation in
LINQ. We show via translations from Eff to Quot and back that the two
approaches are equivalent in expressiveness. Based on the translation
from Eff to Quot, we extend a simple Links compiler to handle queries.
\end{abstract}




\section{Introduction}

Increasingly, programming involves coordinating data and computation
among several layers, such as server-side, client-side and database
layers of a typical three-tier Web application.  The interaction
between the host programming language (e.g. Java, C\#, F\#, Haskell or
some other general-purpose language) running on the server and the
query language (e.g. SQL) running on the database is particularly
important, because the relational model and query language provided by
the database differ from the data structures of most host languages.
Conventional approaches to embedding database queries within a
general-purpose language, such as Java's JDBC, provide the programmer
with precise control over performance but are subject to typing
errors and security vulnerabilities such as SQL injection
attacks~\cite{wassermann07tosem}.  Object-relational mapping (ORM)
tools and libraries, such as Java's Hibernate, provide a popular
alternative by wrapping database access and update in type-safe
object-oriented interfaces, but this leads to a loss of control over
the structure of generated queries, which makes it difficult to
understand and improve performance~\cite{goldschmidt08icse}.  

To avoid these so-called \emph{impedance mismatch} problems, a number
of \emph{language-integrated query} techniques for embedding queries
into general-purpose programming languages have emerged, which seek to
reconcile the goals of type-safety and programmer control.  Two
distinctive styles of language-integrated query have emerged:
\begin{itemize}
\item Employ some form of static
  analysis or type system to identify parts of programs that can be
  turned into queries  (e.g. Kleisli~\cite{Won00}, Links~\cite{CLWY06}, Batches for
  Java~\cite{wiedermann07popl}).
\item Extend a conventional language with explicit facilities for
  quotation or manipulation of query code (e.g. LINQ~\cite{meijer:sigmod}, Ur/Web~\cite{chlipala10pldi},
  Database-Supported Haskell~\cite{dsh}).
\end{itemize}

\begin{figure}

\[
\small
\begin{tabular}{l}
  \begin{tabular}{l}
\fld{employees}\\
\begin{round}
\node (table) [inner sep=0pt] {
\begin{tabular}{l|l|l}
 \fld{dpt} & \fld{name} & \fld{salary} \\ 
\hline
 \str{Product}  & \str{Alex} & 40,000\\  
 \str{Product}  & \str{Bert} & 60,000\\  
 \str{Research} & \str{Cora} & 50,000\\  
 \str{Research} & \str{Drew} &70,000\\  
 \str{Sales}    & \str{Erik} & 200,000\\ 
 \str{Sales}    & \str{Fred} & 95,000\\  
 \str{Sales}    & \str{Gina} & 155,000\\  
\end{tabular}
};
\end{round}
\end{tabular}
\begin{tabular}{l}
\fld{tasks}\\
\begin{round}
\node (table) [inner sep=0pt] {
\begin{tabular}{l|l}
 \fld{emp} & \fld{tsk} \\
\hline
 \str{Alex} & \str{build} \\
 \str{Bert} & \str{build} \\
 \str{Cora} & \str{abstract} \\
 \str{Cora} & \str{build} \\
 \str{Cora} & \str{call} \\
 \str{Cora} & \str{dissemble} \\
 \str{Cora} & \str{enthuse} \\
 \str{Drew} & \str{abstract} \\
 \str{Drew} & \str{enthuse} \\
 \str{Erik} & \str{call} \\
 \str{Erik} & \str{enthuse} \\
 \str{Fred} & \str{call} \\
 \str{Gina} & \str{call} \\
 \str{Gina} & \str{dissemble} \\
\end{tabular}
};
\end{round} 
\end{tabular}
\end{tabular}
\]
\caption{Sample Data}\label{fig:example}
\end{figure}

Links is an example of the first approach.  It uses a type-and-effect
system~\cite{talpin-jouvelot:effect-discipline} to classify parts of
programs as executable only on the database, executable only on the
host programming language, or executable anywhere.  For example,
consider the employee and task data in tables in
Figure~\ref{fig:example}.  The following code
\begin{verbatim}
for (x <- employees)
where(x.salary > 50000)
[(name=x.name)]
\end{verbatim}
retrieves the names of employees earning over \$50,000, specifically
$\collection{\str{Bert}, \str{Drew}, \str{Erik}, \str{Fred},
\str{Gina}}$.  In Links, the same code can be run either on the
database (if \texttt{employees} and \texttt{tasks} are  tables) or in the host
language.  If executed as a query, the interpreter generates a single
(statically defined) SQL query that can take advantage of the database's indexing or
other query optimisation; if executed
in-memory, the expression will by default be interpreted as a
quadratic nested loop.  (Efficient in-memory implementations of query
expressions are also possible~\cite{henglein10hosc}.)

In contrast, in Microsoft's LINQ (supported in C\#, F\#, and some
other .NET languages), the programming language is extended with
query-like syntax.  For
example, the same query as above can be written in F\# as:
\begin{verbatim}
query { for x in employees
        where (x.salary > 50000)
        yield {name=x.name} }
\end{verbatim}
This is just syntactic sugar for code that builds and manipulates
quotations.  In F\#, this facility is built explicitly on top of
language support for quotation~\cite{syme06,fsharp-query-expressions}
and its \emph{computation expression} syntax~\cite{PetricekS14}.
The above F\# query expression is implemented by quoting the code
inside the \verb|query{ ... }| brackets and translating it (at run
time) to C\# values of type \verb|Expression<T>|, which are converted
to SQL by the .NET LINQ to SQL library.

The above example is rather simplistic: the query is \emph{static},
that is, does not depend on any run-time data.  Static queries can be
handled easily even by libraries such as JDBC, and systems such as
Links and LINQ provide the added benefit of type-safety.  However,
most queries are generated \emph{dynamically}, depending on some
run-time data.  The ability to generate dynamic queries is essential
for database programming. Libraries such as JDBC allow queries to be
parameterized over base type values such as strings or integers,
ensuring that values are correctly escaped to prevent SQL injection
attacks.  Both Links and LINQ go significantly further: they allow
constructing dynamic queries using $\lambda$-abstraction and run-time
normalisation, while retaining type safety and preventing SQL
injection.  However, this capability comes with its own pitfalls: it
can be difficult to predict when an expression can be turned into a
single query.

To address this problem, Cooper~\cite{Cooper09} showed how to extend
Links so that performance-critical code can be highlighted with the
\verb|query| keyword.  Links will statically check that the enclosed
expression will definitely translate to a single query (neither
failing at run-time, nor generating multiple queries).  We refer to
this as the \emph{single-query guarantee}. In database theory,
conservativity results due to Wong~\cite{wong96jcss} and others
provided a single-query guarantee in the case of first-order queries:
any query expression having flat input and output types can be turned
into an SQL query. This idea provided the basis for the Kleisli
system~\cite{Won00}, which was a source of inspiration to Links; the single-query
guarantee was generalised to the higher-order case by
Cooper~\cite{Cooper09}, who also gave a static type-and-effect system
that showed how to embed queries in a higher-order general-purpose
language.  Subsequent work on Links~\cite{lindley12tldi} generalised
this to use row typing and effect polymorphism.

The possibility of generating LINQ queries dynamically in ad hoc cases
was discussed by Syme~\cite{syme06} and
Petricek~\cite{petricek-dynamic-linq,petricek-dynamic-flinq}. The F\#
and LINQ to SQL libraries in Microsoft .NET do not provide a
single-query guarantee for dynamic queries; instead, they attempt to
generate a single query but sometimes fail or generate multiple
queries.  In our recent paper~\cite{cheney13icfp} we showed that
Cooper's approach to normalisation for Links can be transferred to
provide systematic support for abstraction in LINQ in F\#, providing a
single-query guarantee.  In the rest of this paper, we consider the
F\# LINQ approach with this extension.

Nevertheless, there are still apparent differences between the
approaches.  For example, in LINQ, a query expression cannot be
(easily) reused as ordinary code.  This potentially leads to the
need to write (essentially) the same code twice, once for ordinary use
and once for use on the database.  Code duplication can interfere with
the use of functional
abstraction to construct queries.  For example, the
following Links code
\begin{verbatim}
fun elem(x,xs) {
  not(empty(for (y <- xs) where (x == y) [()]))
}
fun canDo(name,tsk) {
  elem("build", for (t <- tasks)
                   where (t.emp == name)
                   [t.tsk])
}
query { for (x <- employees)
        where (canDo(x.name,"build"))
        [(name=x.name)] }
\end{verbatim}
defines functions \verb|elem| and \verb|canDo| that test respectively
whether a value is an element of a collection and whether an employee
can do a certain task. The Links effect system correctly determines
that \verb|elem| can be run anywhere, and that \verb|canDo| can be run
on the database. When the query is to be executed, Links normalises
the query by inlining \verb|elem| and \verb|canDo| and performing
other transformations to generate a single SQL query~\cite{Cooper09}.
In contrast, naively executing this code might involve loading all of
the data from the \verb|employees| table, and running one subquery to
compute \verb|canDo| for each \verb|employees| row in-memory.

In F\#, it is possible to do something similar, but only by explicitly
quoting \verb|elem| and \verb|canDo|.
\begin{verbatim}
let elem = <@ fun x xs ->
                query { for y in xs
                        exists(y = x) } @>
let canDo = <@ fun name tsk ->
                 (%elem) tsk (for t in tasks
                              where (t.emp = name)
                              yield t.tsk) @>
query { for x in employees
        where ((%canDo) x.name "build")
        yield {name=x.name} }
\end{verbatim}
The quoted version of \verb|elem| is spliced into the query using
antiquotation \verb|(%elem)|. If we need the
\verb|elem| function in both query and non-query code, its code must
be duplicated, or we need to evaluate or generate compiled code for it
at runtime.  (F\#'s quotation library does include \verb|Eval| and
\verb|Compile| functions that can be used for this purpose, but it is
not clear that these actually generate efficient code at runtime, nor
is it convenient to write this boilerplate code.)

It is important to note that SQL does not natively support general
recursion or first-class functional abstraction (although there are
recent proposals to support the
latter~\cite{grust13dbpl}). Nonrecursive lambda-abstraction is
supported in query expressions in Links and F\#, but it is eliminated
in the process of generating an SQL query.  Recursive functions can
also be used to construct queries from data in the host language, in
both Links and LINQ, but care is needed to make the staging explicit.
For example, in F\# we can define a predicate that tests whether an
employee can do all tasks in some list as follows:
\begin{verbatim}
let rec canDoAll(tsks) = 
  match tsks with
    [] -> <@ fun name -> true @>
  | tsk::tsks' -> <@ fun name -> 
      (%canDo) name tsk && (%canDoAll tsks') name @>
query {
  for x in employees
  where ((%canDoAll ["build","call"]) x.name)
  yield {name=x.name} }
\end{verbatim}
This is also possible in Links, but we need to use function
abstraction and hoist subcomputations to satisfy the effect type
system:
\begin{verbatim}
fun canDoAll(tsks) {
  switch (tsks) {
    case [] -> fun (name) {true}
    case (tsk::tsks') -> 
      var p = canDoAll(tsks');
      fun (name) { canDo(name,tsk) && p(name) } } }
query {
  for (x <- employees)
  where (canDoAll(["build","call"])(x.name))
  [(name=x.name)] }
\end{verbatim}
We have to hoist the recursive call to \verb|canDoAll| and name it so
that it is clear to the type system that the recursive computation
does not depend on values in the database that are not directly
available to the host language interpreter.  Arguably, in this case
F\#'s explicit quotation and antiquotation annotations clarify the
distinction between staging and functional abstraction, whereas in
Links the distinction is not as explicit in the program syntax.

Both techniques are essentially heterogeneous forms of staged
computation, based on a common foundation of manipulating
partially-evaluated query expressions (or query fragments) at run time
in order to construct SQL queries.  The single-query
property~\cite{Cooper09,cheney13icfp} guarantees each query expression
succeeds in generating one and only one query, even if
lambda-abstraction or recursion is used to construct the queries.
However, several natural questions about the relative strengths of the
two approaches remain unanswered: Can we translate the Links
effect-based approach to the (seemingly lower-level) LINQ
quotation-based approach?  If so, this might suggest a fruitful
implementation strategy.  Conversely, do we lose any expressiveness by
providing the (seemingly higher-level) effect-based Links approach?
Or can we always (in principle) translate LINQ-style quotation-based
code to Links-style effect-based code?

In this paper, we consider the problem of relating the
\emph{expressiveness} of the two approaches to language-integrated
query represented by Links and LINQ.  Database query languages are
often limited in expressiveness: for example, plain SQL conjunctive
queries cannot express recursive properties such as transitive
closure.  Thus, understanding the relative expressiveness of different
(Turing-incomplete) query languages, and the tradeoff with complexity
of query evaluation or optimisation, are important issues in database
theory~\cite{ahv}.  We are interested in a dual question: what is the
expressiveness of a programming language that generates queries?
Given that both Links and LINQ approaches provide a measure of support
for dynamic queries, can they express the same classes of dynamic
queries?

To make this question precise, we introduce two core calculi: $\EffSub$,
representing the effect-based approach supported by Links, and
$\Quot$, representing the quotation-based approach adopted in LINQ in
F\#.  The former is similar to Cooper's core language~\cite{Cooper09};
the latter is essentially the same as the T-LINQ core
language~\cite{cheney13icfp}.  Both core languages make simplifying
assumptions compared to Links and F\# respectively, but we argue that
they capture what is essential about the two approaches, making them
suitable for a formal comparison that avoids preoccupation with other
distracting details (e.g. Links's support for client-side
programming~\cite{CooperW09} or F\#'s support for
objects~\cite{fsharp3}, or different facilities for polymorphism in
both languages.)
 
In database theory, the expressiveness of a language is usually
measured by the set of functions definable in it, according to a
conventional denotational semantics of database queries. However, this
notion of expressiveness is not very interesting for general-purpose
languages: two Turing-complete programming languages are always (by
definition) expressively equivalent in this sense.
Felleisen~\cite{felleisen91scp} and Mitchell~\cite{mitchell93scp}
proposed notions of expressiveness based on restricted forms of
translation among different languages.  However, neither of these
notions seems appropriate for relating language-integrated query
formalisms.

Programs interact with a database that may be concurrently
updated by other programs, and we want a notion of equivalence that
takes query behaviour into account while abstracting over the possible
concurrent behaviours of the database.  For example, we want to
consider a program that issues query $Q$ to the database
\emph{inequivalent} to another program that reads all the database
tables into memory and executes $Q$ in-memory.  In addition, we wish
to abstract as much as possible over the possible behaviours of the
database: databases are typically concurrently accessed and updated by
many applications, and we want our notion of expressiveness to minimise
assumptions about the behaviour of the database.  Thus, we define the
semantics of both languages as labeled transitions, where labels are
either silent transitions or pairs $(q,V)$ consisting of
database queries and responses.  We consider two programs
\emph{query-equivalent} if, given the same input, they have the same
possible (finite and infinite) query/response traces
$(q_1,V_1),\ldots,(q_n,V_n),\ldots$.

We show that $\EffSub$ programs can be translated to query-equivalent
$\Quot$ programs via a two-stage translation: first we eliminate
subeffecting by duplicating code in the \emph{doubling} translation,
then we introduce explicit quotation and antiquotation in the
\emph{splicing} translation. Perhaps more surprisingly, we can also give a
converse translation from $\Quot$ to $\EffSub$, which translates quoted
code to thunks (functions with unit domain): thus, the two approaches
are expressively equivalent up to query-equivalence.

The current version of Links is interpreted, and query normalisation
depends on being able to inspect code at run time. The translation
from $\EffSub$ to $\Quot$ suggests a compilation strategy by translating
Links-style code to explicitly quoted code.

In the rest of this paper, we present the following contributions:
\begin{itemize}
\item We propose an appropriate notion of dynamic query behaviour
  suitable for comparing the expressiveness of different
  language-integrated query techniques.
\item We provide a detailed exploration of the relationship between
  implicit, effect-based (Links/$\EffSub$) and explicit,
  quotation-based (LINQ/$\Quot$) approaches, giving type- and
  semantics-preserving translations in each direction.
\item We discuss an application of the translation from $\EffSub$ to
  $\Quot$ to support compilation of Links programs with embedded
  queries, along with preliminary experimental results.
\end{itemize}

The rest of this paper is structured as follows.
Section~\ref{sec:background} presents necessary background material
from prior work, and defines the desired notion of equivalence of
programs with respect to observable query behaviour.
Section~\ref{sec:formalisation} presents 
$\EffSub$ and $\Quot$, giving their syntax, type systems, and
operational semantics.
Section~\ref{sec:translations} presents translations between $\EffSub$
and $\Quot$.  
Section~\ref{sec:implementation} presents a practical application of
the translation from $\EffSub$ to $\Quot$, which serves as the basis
for a prototype compiler for Links that supports run-time dynamic
query generation.
Section~\ref{sec:related} provide additional
discussion of related work and Section~\ref{sec:concl} concludes.

\section{Background}
\label{sec:background}

The Nested Relational Calculus (NRC) is a widely-studied core language
for database queries corresponding closely to monadic comprehension
syntax~\cite{BNTW95,buneman+:comprehensions}.  Previous
work~\cite{wong96jcss,Cooper09,lindley12tldi,cheney13icfp} has shown
how first- and higher-order variants of NRC can be used for
language-integrated query.  We give the syntax of first-order NRC in
Figure~\ref{fig:nrc}.  Extended examples of the use of NRC are
presented in prior work~\cite{buneman+:comprehensions,cheney13icfp}.

We let $x$ range over variables, $c$ range over constants, and $\op$
range over primitive operators. Records $\record{\overline{\ell=q}}$
and field projections $q.\ell$ are standard. We write $\zero$ for an
empty bag, $\singleton q$ for the singleton bag containing the element
$q$, and $q \plus q'$ for the union of bags $q$ and $q'$. We write
$\For\,(x^A \gets q)~q'$ for a bag comprehension, which for each
element $x$ in $q$ evaluates $q'$, then computes the union of the
resulting bags. We write $\Table~t$ for the relational database table
$t$. In order to keep normalisation as simple as possible we restrict
ourselves to one sided conditionals over collections (equivalent to
SQL where clauses). The expression $\If~q~q'$ evaluates to $q'$ if $q$
evaluates to $\True$ and $\zero$ if $q$ evaluates to $\False$. Lindley
and Cheney~\cite{lindley12tldi} describe how to normalise in the
presence of general conditionals; briefly, the idea is to push
conditionals inside records and translate
$\If~q~q'~q''$ to $(\If~q~q')
\plus (\If~(\neg q)~q'')$ when $q',q''$ are of bag type.

The NRC types include base types ($\Int$, $\Bool$, $\String$, etc.),
record types $\record{\overline{\ell:A}}$, and bag types
$\tColl{A}$. Row types are \emph{flat} record types restricted to
contain base types (just like rows in SQL queries).

Conservativity results (see e.g. Wong~\cite{wong96jcss}) ensure that
any NRC expression $M$ having a flat return type and flat inputs can
be normalised to a form that corresponds directly to SQL.


We assume a fixed signature $\Sigma$ mapping constants $c$ to base
types, operators $op$ to functions on base types, and table references
to flat bag types $\collection{{R}}$. We omit typing or evaluation
rules for queries; these are standard and implicit in the typing and
operational semantics rules of $\EffSub$ and $\Quot$ given later.

We will model the behaviour of a database server nondeterministically:
whenever a query is posed, the response may be any value of the
appropriate type.  Thus, we fix a set $\Omega$ of all pairs 
$(q,V)$ such that whenever $\vdash q : A$ we have $\vdash V : A$.
Further constraints, reflecting the semantics of the query language or
integrity constraints on the database tables, could be imposed.  Our
results concerning expressiveness are parametric in $\Omega$ (provided
it is at least type-safe  and respects query equivalence).

\begin{figure}
\begin{mysyntax}
\cat{Query}  &q  &::=& x \mid c \mid \op(\overline{q}) \mid \If~q~q'
     \mid\record{\overline{\ell=q}} \mid q.\ell \\
     &&\mid& \zero  \mid \singleton{q} \mid q_1 \plus q_2 \mid \For\,(x \gets q)\, q'\\
     &&\mid& \Table~t \\
     \cat{Base type} &O&::=& \Int \mid \Bool \mid \String \mid \cdots\\
\cat{Type} &A,B&::=& O \mid \record{\overline{\ell=A}} \mid \tColl{A}\\
  \cat{Row type} & R &::=& \record{\overline{\ell:O}} 
\end{mysyntax}%
\caption{Syntax of NRC}
\label{fig:nrc}
\end{figure}

\begin{definition}
  Let $L$ be a set of actions, including a ``silent'' action $\tau$,
  and let $\mu$ range over elements of $L$. A labeled transition
  system over some set of labels $L$ is a structure $(X,\too)$ where
  ${\too} \subseteq X \times L \times X$.  We write $x \ltoo{\mu} y $
  when $(x,\mu,y) \in {\too}$ for some $\mu \in L$ and $x \too y$ when
  $x \ltoo{\tau} y$.  We write $\Too$ for $\too^*$ and write
  $\lToo{\mu}$ for $\Too \circ \ltoo{\mu} \circ \Too$. Suppose
  that $x_1 \lToo{\mu_1} x_2 \lToo{\mu_2} \cdots x_n
  \lToo{\mu_n} \cdots$ is a (finite or infinite) run starting from
  $x_1$.  We say that the (finite or infinite) word $\mu_1 \cdots
  \mu_n \cdots$ obtained from concatenating the non-$\tau$ labels
  is the \emph{trace} of the run, and we define $Tr(x)$ to be the set
  of all traces starting from $x$.
  \end{definition}
  \begin{definition}
    Let $(X,\too_X)$ and $(Y,\too_Y)$ be labeled transition systems over
    the same label set $L$.  We say that $x \in X$
    \emph{trace-simulates} $y \in Y$ (or $x \trsim y$) if $Tr(x)
    \subseteq Tr(y)$, and $x \in X$ and $y \in Y$ are
    \emph{trace-equivalent} ($x \trbisim y$) if $Tr(x) = Tr(y)$.
\end{definition}
In the rest of this paper, we are concerned with instances of
trace-equivalence where the labels are either $\tau$ or pairs $(q,V)
\in \Omega$.

\section{Formalisation}
\label{sec:formalisation}
\subsection{Source Language: $\EffSub$}

\begin{figure*}

\begin{smathpar}
  \boxed{\Gamma \vdash M : A \eff E}\qquad \boxed{E_1 \leq E_2} \hfill \\
  \inferrule[Var] { } {\Gamma, x:A \vdash x : A \eff E}
\and
  \inferrule[Const] {\Sigma(c) = A} {\Gamma \vdash c : A \eff E}
\and
  \inferrule[Op]
  {\Sigma(\op) = (\overline{O}) \to O \\
    \overline{\Gamma \vdash M : O \eff E}} {\Gamma \vdash
    \op(\overline{M}) : O \eff E}
\and
  \inferrule[Lam] {\Gamma, x:A \vdash M : B \eff E} {\Gamma \vdash
    \lambda^E x^A.M:A \TO{E} B \eff E'}
\and
  \inferrule[App] {\Gamma \vdash M : A \TO{E'} B \eff E \\ \Gamma
    \vdash N : A \eff E \\ E' \leq E} {\Gamma \vdash M~N : B \eff E}
\and
  \inferrule[If]
  {\Gamma \vdash L : \Bool \eff E\\
    \Gamma \vdash M : \tColl{A} \eff E} {\Gamma \vdash \If\,L\,M : \tColl{A} \eff
    E}
\and
  \inferrule[Record] {\overline{\Gamma \vdash M : A \eff E}} {\Gamma
    \vdash \record{\overline{\ell=M}} : \record{\overline{\ell:A}} \eff E}
\and
  \inferrule[Project] {\Gamma \vdash M : \record{\overline{\ell:A}}
    \eff E} {\Gamma \vdash M.{\ell_i} : A_i \eff E}
\and
  \inferrule[Empty] { } {\Gamma \vdash \zero : \tColl{A} \eff E}
\and
  \inferrule[Singleton] {\Gamma \vdash M : A \eff E} {\Gamma \vdash
    \singleton{M} : \tColl{A} \eff E}
\and
  \inferrule[Union] {\Gamma \vdash M : \tColl{A} \eff E \\ \Gamma \vdash
    N : \tColl{A} \eff E} {\Gamma \vdash M \plus N : \tColl{A} \eff E}
\and
  \inferrule[For]
  {\Gamma \vdash M : \tColl{A} \eff E \\
    \Gamma,x:A \vdash N : \tColl{B} \eff E} {\Gamma \vdash \For\,(x^A
    \gets M)\,N : \tColl{B} \eff E}
\and
  \inferrule[Table] {\Sigma(t) = \tColl{A}} {\Gamma \vdash \Table~t :
    \tColl{A} \eff \db}
\and
  \inferrule[Query] {\Gamma \vdash M : \tColl{R} \eff \db} {\Gamma
    \vdash \Query~M : \tColl{R} \eff \pl}
\and
%
%
  \inferrule[Rec] {\Gamma,f:A \TO{\pl} B,x:A \vdash M : B \eff \pl}
  {\Gamma \vdash \rec{f^{A \TO{\pl} B}}{x^A}{M} : A \TO{\pl} B \eff E}
\and
  \inferrule[Fold]
  {\Gamma \vdash L : \tColl{A} \eff \pl \\
    \Gamma \vdash M : B \eff \pl \\
    \Gamma \vdash N : A \TO{\pl} B \TO{\pl} B \eff \pl} {\Gamma \vdash
    \fold~L~M~N : B \eff \pl}
\and
  \inferrule[Reflexivity] { } {E \leq E}
\and
  \inferrule[AnyPl] { } {\any \leq \pl}
\and
  \inferrule[AnyDb] { } {\any \leq \db}
\end{smathpar}
\caption{\label{fig:eff-typing}Typing and Subeffecting Rules for
  $\EffSub$}
\end{figure*}
\begin{figure*}

\begin{smathpar}
  \boxed{\Gamma \vdash M : A}\hfill \\

  \inferrule[Var] { } {\Gamma, x:A \vdash x : A}
\and
  \inferrule[Const] {\Sigma(c) = A} {\Gamma \vdash c : A}
\and
  \inferrule[Op]
  {\Sigma(\op) = (\overline{O}) \to O \\
    \overline{\Gamma \vdash M_i : O_i}} {\Gamma \vdash \op(\overline{M}) : O}
\and
  \inferrule[Lift] {\Gamma \vdash M : O} {\Gamma \vdash
    \lift{M} : \tQuote{O}}
\and
  \inferrule[Lam] {\Gamma, x:A \vdash M : B} {\Gamma \vdash \lambda
    x^A.M:A \to B}
\and
  \inferrule[App] {\Gamma \vdash M : A \to B \\ \Gamma \vdash N : A}
  {\Gamma \vdash M~N : B}
\and
  \inferrule[If]
  {\Gamma \vdash L : \Bool \\
    \Gamma \vdash M : \tColl{A}} {\Gamma \vdash \If\,L\,M : \tColl{A}}
\and
  \inferrule[Record] {\overline{\Gamma \vdash M : A}} {\Gamma \vdash
    \record{\overline{\ell=M}} : \record{\overline{\ell:A}}}
\and
  \inferrule[Project] {\Gamma \vdash M : \record{\overline{\ell:A}}}
  {\Gamma \vdash M.{\ell_i} : A_i}
\and
  \inferrule[Empty] { } {\Gamma \vdash \zero : \tColl{A}}
\and
  \inferrule[Singleton] {\Gamma \vdash M : A} {\Gamma \vdash
    \singleton{M} : \tColl{A}}
\and
  \inferrule[Union] {\Gamma \vdash M : \tColl{A} \\ \Gamma \vdash N :
    \tColl{A}} {\Gamma \vdash M \plus N : \tColl{A}}
\and
  \inferrule[For]
  {\Gamma \vdash M : \tColl{A} \\
    \Gamma,x:A \vdash N : \tColl{B}} {\Gamma \vdash \For\,(x \gets M)\,N
    : \tColl{B}}
\and
  \inferrule[Quote] {\Gamma; \cdot \vdash M : A} {\Gamma \vdash
    \Quote{M} : \tQuote{A}}
\and
  \inferrule[Query] {\Gamma \vdash M : \tQuote{\tColl{R}}}
  {\Gamma \vdash \Query~M : \tColl{R}}
\and
  \inferrule[Rec] {\Gamma,f:A \to B,x:A \vdash M : B} {\Gamma \vdash
    \rec{f^{A \to B}}{x^A}{M} : A \to B}
\and
  \inferrule[Fold]
  {\Gamma \vdash L : \tColl{A} \\
    \Gamma \vdash M : B \\
    \Gamma \vdash N : A \to B \to B} {\Gamma \vdash \fold~L~M~N : B}
  \\
  \boxed{\Gamma; \Delta \vdash M : A} \hfill \\
  \inferrule[VarQ] { } {\Gamma; \Delta, x:A \vdash x : A}
\and
  \inferrule[ConstQ] {\Sigma(c) = A} {\Gamma; \Delta \vdash c : A}
\and
  \inferrule[OpQ]
  {\Sigma(\op) = (\overline{O}) \to O \\
    \overline{\Gamma; \Delta \vdash M : O}} {\Gamma; \Delta \vdash
    \op(\overline{M}) : O}
\and
  \inferrule[LamQ] {\Gamma; \Delta, x:A \vdash M : B} {\Gamma; \Delta
    \vdash \lambda x^A.M:A \to B}
\and
  \inferrule[AppQ] {\Gamma; \Delta \vdash M : A \to B \\ \Gamma;
    \Delta \vdash N : A} {\Gamma; \Delta \vdash M~N : B}
\and
  \inferrule[IfQ]
  {\Gamma; \Delta \vdash L : \Bool \\
    \Gamma; \Delta \vdash M : A} {\Gamma; \Delta \vdash \If\,L\,M :
    A}
\and
  \inferrule[RecordQ] {\overline{\Gamma; \Delta \vdash M : A}}
  {\Gamma;\Delta \vdash \record{\overline{\ell=M}} :
    \record{\overline{\ell:A}}}
\and
  \inferrule[ProjectQ] {\Gamma; \Delta \vdash M :
    \record{\overline{\ell:A}}} {\Gamma; \Delta \vdash M.{\ell_i} :
    A_i}
\and
  \inferrule[EmptyQ] { } {\Gamma; \Delta \vdash \zero^A : \tColl{A}}
\and
  \inferrule[SingletonQ] {\Gamma; \Delta \vdash M : A} {\Gamma; \Delta
    \vdash \singleton{M} : \tColl{A}}
\and
  \inferrule[UnionQ] {\Gamma;\Delta \vdash M : \tColl{A} \\ \Gamma;\Delta \vdash N :
    \tColl{A}} {\Gamma;\Delta \vdash M \plus N : \tColl{A}}
\and
  \inferrule[ForQ]
  {\Gamma; \Delta \vdash M : \tColl{A} \\
    \Gamma; \Delta, x:A \vdash N : \tColl{B}} {\Gamma; \Delta \vdash
    \For\,(x \gets M)\,N : \tColl{B}}
\and
  \inferrule[Table] {\Sigma(t) = \tColl{R}} {\Gamma; \Delta \vdash
    \Table~t : \tColl{R}}
\and
  \inferrule[Antiquote] {\Gamma \vdash M : \tQuote{A}}
  {\Gamma; \Delta \vdash \AntiQuote{M} : A}
\end{smathpar}
\caption{\label{fig:quot-typing}Typing Rules for $\Quot$}
\end{figure*}

$\EffSub$ is a higher-order nested relational calculus over bags
augmented with an effect type system for issuing flat NRC relational
queries.
It is similar to calculi considered in previous work on query
compilation~\cite{Cooper09,lindley12tldi}. A difference is that
$\EffSub$ models two ground effects ($\pl$ and $\db$), whereas the
other calculi model just one (indicating code that cannot be run in
the database). Like Cooper's system, $\EffSub$ provides
subeffecting. Our previous work~\cite{lindley12tldi}, and its
implementation in the Links web programming language~\cite{CLWY06}
uses row-based effect polymorphism instead; while effect polymorphism
is more flexible, we focus on the simpler subtype-based approach here.

The effects of $\EffSub$ are given by the following grammar.
\begin{mysyntax}
  \cat{ground effects} &X, Y &::=& \pl \mid \db \\
  \cat{effects}        &   E &::=& X \mid \any \\
\end{mysyntax}%
Code that runs in the programming language has the $\pl$ effect. Code
that runs in the database has the $\db$ effect. Code that can be run
in either place, i.e. anywhere, has the $\any$ effect. The typing
rules will enforce a \emph{subeffecting} discipline, allowing a
function that has the $\any$ effect to be applied with either the
$\pl$ or the $\db$ effect.

The types of $\EffSub$ are given by the following grammar.
\begin{mysyntax}
  \cat{types} & A, B &::=& O \mid
  \record{\overline{\ell:A}}
  \mid \tColl{A} \mid A \TO{E} B 
\end{mysyntax}%
Types in $\EffSub$ extend the NRC base types, records, and collections
with function types $A
\TO{E} B$ annotated by an effect $E$.

The terms of $\EffSub$ are given by the following grammar.
\begin{equations}
  L,M,N &::=& x \mid c \mid \op(\overline{M}) \mid \lambda^E x^A.M \mid M~N \\
  &\mid& \If~L~M \mid \record{\overline{\ell=M}} \mid M.\ell \\
  &\mid& \zero \mid \singleton M  \mid M \plus N 
 \mid \For\,(x^A \gets M)~N \\
  &\mid& \rec{f^{A \TO{\pl} B}}{x^A}{M} \mid \fold\,L\,M\,N \\
  &\mid& \Query\,M \mid \Table~t 
\end{equations}%
We will often omit type and effect annotations on bindings. We include
them so that later we can define translations on terms rather than on
judgements (Section~\ref{sec:translations}).


The syntax is in most cases similar to NRC or standard.
Lambda-abstractions $\lambda^E x^A.M$ are annotated with an effect
$E$. Recursive functions $\rec{f}{x}{M}$ and folds are only available
in the programming language.  The fold operation is the only
elimination form for bag values.  The special form $\Query\,M$ denotes
a query returning results of type $\tColl{R}$, where $M$ has type
$\tColl{R}$.

A type environment ascribes types to variables.
\[
\Gamma, \Delta ::= \cdot \mid \Gamma, x:A
\]

The typing rules are given in Figure~\ref{fig:eff-typing}. In this
figure and elsewhere, notations such as $\overline{M}$ or
$\overline{\Gamma \vdash M : A \eff E}$ abbreviate lists of terms,
judgments, etc. The typing judgment is of the form $\Gamma \vdash M :
A \eff E$, indicating that in context $\Gamma$, term $M$ has type $A$
and effect $E$.  The effect $E$ can be either $\pl$ or $\db$,
indicating that $M$ can only be executed by the host programming
language or by the database respectively, or it can be $\any$,
indicating that $M$ can be executed in either context.

We again assume a signature $\Sigma$ that maps each constant $c$ to
its underlying type, each primitive operator $\op$ to its type
(e.g. $\Sigma(\wedge) = (\Bool,\Bool) \to \Bool$ and $\Sigma(+) =
(\Int,\Int) \to \Int$), and each table $t$ to the type of its rows.
(For simplicity we assume that the same base types, constants and
primitive operations are available to both the host and database.)
%
The rules are quite standard; most of them are parametric in the
effect.
The most interesting rules are those that do something non-trivial
with effects, such as \textsc{Table}, \textsc{Query}, \textsc{Rec},
\textsc{Fold} and \textsc{App}. The \textsc{Table} rule ensures that
table references $\Table~t$ can only be used in a database
context. The \textsc{Query} rule requires that the body of a query
expression $\Query~M$ must be run in the database, and the query
expression itself must be invoked from the programming language. This
rule also implicitly requires that a query result type $\tColl{R}$ is
a bag of records (note that $R$ is a row type).  Recursive functions
(rule \textsc{Rec}) and folds (rule \textsc{Fold}) can only be applied
in the programming language. The \textsc{App} rule allows a function
with the $\any$ effect to be applied in the programming language or
the database.


%

We define a sub-language $\Eff$ by restriction of $\EffSub$ to
programs in which the $\any$ effect is disallowed. The subeffecting
constraint in the \textsc{App} rule is superfluous in this
sublanguage.
%
We can translate any $\EffSub$ program to an equivalent, albeit
longer, $\Eff$ program, by representing each $\any$ function as a pair
of a $\pl$ function and a $\db$ function
(Section~\ref{sec:effsub-to-eff}).

\subsection{Operational Semantics for $\EffSub$}

We now present small-step operational semantics for $\EffSub$.
The syntax of values and evaluation contexts is given in
Figure~\ref{fig:eff-values}.
The values are standard. We write $\collection{\overline{V}}$ for
$\singleton{V_1} \plus \dots \plus \singleton{V_n} \plus \zero$.
The operational semantics is defined by reduction relation $M
\ltoo{\mu} N$ as shown in
Figure~\ref{fig:eff-semantics}.  It is
parameterised by an interpretation $\delta$ for each primitive
operation $\op$, and a set $\Omega$ of possible query request and
response pairs $(q,V)$, both of which respect types: if $\Sigma(\op) =
\overline{O} \to O$ and $\vdash \overline{V:O}$ and $V =
\delta(\op,\overline{V})$ then $\vdash V:O$, and if $(q,V) \in \Omega$
and $\vdash q : \tColl{R}$ then $\vdash V : \tColl{R}$.



%

\begin{figure}
\begin{mysyntax}
\multicolumn{4}{@{}l@{}}{\cat{value}} \\
 & V,W &::=& c
    \mid \lambda x.M
    \mid \rec{f}{x}{M}
    \mid \record{\overline{\ell=V}}
    \mid \collection{\overline{V}} \\
\multicolumn{4}{@{}l@{}}{\cat{evaluation context}} \\
& \EC &::=& [~]
    \mid \op(\overline{V},\EC,\overline{M})
    \mid \EC~ M \mid V~ \EC \\
 && \mid & \record{\overline{\ell=V},\ell'=\EC,\overline{\ell''=M}}
    \mid \EC.\ell
    \mid \singleton{\EC} \\
 && \mid & \EC \plus M
    \mid \singleton{V} \plus \EC
    \mid \For\,(x \gets \EC)\, N \\
 && \mid & \If~\EC~M
\end{mysyntax}%

\caption{Values and Evaluation Contexts for $\Eff$}
\label{fig:eff-values}


\begin{equations}
\op(\overline{V}) &\evalto& \opdelta(\op, \overline{V}) \\
 (\lambda x.M)~ V &\evaltoX& M[x := V] \\
 (\rec{f}{x}{M})~ V &\evaltopl& 
   M[f := \rec{f}{x}{M}, x := V] \\
 \record{\overline{\ell=V}}.\ell_i &\evaltoX& V_i \\
 \If~\True~M  &\evaltoX& M \\
 \If~\False~M &\evaltoX& \zero \\
\zero \plus V &\evaltoX& V\\
(\singleton{V} \plus V') \plus V'' &\evaltoX& \singleton{V} \plus (V'
\plus V'')\\
\For\, (x \gets \zero)\, N &\evaltoX& \zero \\
\For\, (x \gets \singleton{V} \plus W)\, N &\evaltoX&
   N[x:=V]~\plus\\&&\qquad (\For\, (x \gets W)\, N) \\
 \fold\,\zero\,M\,N &\evaltopl& M \\
 \fold\,((\singleton{V}) \plus W)\,M\,N &\evaltopl& N~V\,(\fold\,W\,M\,N) \\
\Query~M &\ltoo{(\normEff{M},V)}& V \hfill ((\normEff{M},V) \in \Omega)\\ \\
\end{equations}%
\vspace{-1em}
\begin{mathpar}
\inferrule
{M \ltoo{\mu} N}{\EC[M] \ltoo{\mu} \EC[N]}
\end{mathpar}%
\caption{Operational Semantics for $\EffSub$}
\label{fig:eff-semantics}
\end{figure}


The rules are standard apart from the
one for query evaluation. Evaluation contexts $\EC$ enforce
left-to-right call-by-value evaluation.  Rule (query) evaluates a
query $M$ by first \emph{normalising} $M$ to yield an equivalent NRC
query $q = \normEff{M}$, and then taking a transition labeled $(q,V)$
yielding result value $V$. The normalisation function $|-|$ is the
same as that of Cheney et al.~\cite{cheney13icfp}. It first applies
standard symbolic reduction rules (Figure~\ref{fig:query-rewriting})
and then further ad hoc reduction rules
(Figure~\ref{fig:adhoc-rewriting}). The former eliminate all nesting
and abstraction from a closed term of flat bag type, while the latter
account for the lack of uniformity in SQL (see \cite{cheney13icfp} for
further details). Define $\normEff{L} = N$ when $L \rewriteto^* M$ and
$M \adhocto^* N$, where $M$ and $N$ are in normal form with respect to
$\rewriteto$ and $\adhocto$ respectively.


\begin{figure}
\begin{equations}
(\lambda x.M)~ N &\rewriteto&  M[x := N] \\
\record{\overline{\ell=M}}.\ell_i &\rewriteto& M_i \\
\For\,(x \gets \singleton M)~N &\rewriteto& N[x := M] \\
\For\,(y \gets \For\,(x \gets L)~M)~N &\rewriteto&
  \For\,(x \gets L)~\For\,(y \gets M)~N \\
\For\,(x \gets \If~L~M)~N &\rewriteto& \If~L~(\For\,(x \gets M)~N) \\
\For\,(x \gets \zero)\,N &\rewriteto& \zero \\
\For\,(x \gets (L \plus M))\,N &\rewriteto& \\
\multicolumn{3}{r}{\quad (\For\,(x \gets L)\,N) \plus (\For\,(x \gets M)\,N)} \\
\If~\True~M &\rewriteto& M \\
\If~\False~M &\rewriteto& \zero \\
\end{equations}%
\caption{Normalisation Stage 1: symbolic reduction}
\label{fig:query-rewriting}

\begin{equations}
\For\,(x \gets L)\,(M \plus N) &\adhocto&
  (\For (x \gets L)\,M) \plus (\For\,(x \gets L)\,N) \\
\For\,(x \gets L)\,\zero &\adhocto& \zero \\
\If~L~(M \plus N) &\adhocto& (\If~L~M) \plus (\If~L~N) \\
\If~L~\zero &\adhocto& \zero \\
\If~L~(\If~M~N) &\adhocto& \If~(L \ampersand M)~N \\
\If~L~(\For\,(x \gets M)\,N) &\adhocto& \For\,(x \gets M)~(\If~L~N) \\
\end{equations}%
\caption{Normalisation Stage 2: ad hoc reduction}
\label{fig:adhoc-rewriting}

\end{figure}

It is straightforward to show type soundness via the usual method of
preservation and progress.
\begin{proposition}
  If\/ $\Gamma \vdash M : A \eff E$ and $M \ltoo{\mu} N$ then $\Gamma
  \vdash N : A \eff E$.  If\/ $\Gamma \vdash M : A \eff E$ then either
  $M$ is a value or $M \ltoo{\mu} N$ for some $N$ and $\mu$.
\end{proposition}

\subsection{Target Language: $\Quot$}

$\Quot$ is a higher-order nested relational calculus over bags
augmented with a quotation mechanism for constructing NRC queries using
quotation.

Modulo superficial differences $\Quot$ is essentially the same as the
\TLINQ core language~\cite{cheney13icfp}. Specifically, the
differences are: the lexical syntax; $\Quot$ includes $\fold$; $\Quot$
includes extra type annotations (to aid the translation to $\Eff$);
\TLINQ has a $\Database$ construct whereas $\Quot$ has a $\Table$
construct.

Where $\EffSub$ uses effect types to distinguish the \emph{programming
  language} from the \emph{database}, $\Quot$ uses quotation to
distinguish the \emph{host language} from the \emph{query
  language}. Host language terms may build and evaluate quoted query
language terms.  For convenience (and ease of comparison with
$\EffSub$), we use the same syntax for the host and query languages,
so that the query language is essentially a sublanguage of the host
language (except for antiquotation and $\Table~t$).  In general, the
two languages may be different, as already discussed
elsewhere~\cite{cheney13icfp}.

The types of $\Quot$ are given by the following grammar.
\begin{mysyntax}
  \cat{types} & A, B &::=& O \mid
  \record{\overline{\ell:A}} \mid \tColl{A}
  \mid A \to B \mid \tQuote{A} \\
\end{mysyntax}%
They are the same as for $\EffSub$, except function types are not
annotated with effects, and types are extended to include
\emph{quotation types} $\tQuote{A}$, which represent closed, quoted
query terms of type $A$.

The terms of $\Quot$ are given by the following grammar.
\begin{equations}
  L,M,N &::=& x \mid c \mid \op(\overline{M}) \mid \lambda x^A.M \mid M~N \\
  &\mid& \If~L~M \mid \record{\overline{\ell=M}} \mid M.\ell \\
  &\mid& \zero \mid\singleton M \mid M \plus N \mid \For\,(x^A \gets M)~N \\
  &\mid& \rec{f^{A \to B}}{x^A}{M} \mid \fold\,L\,M\,N \\
  &\mid& \Query~M \mid \Table~t \\
  &\mid& \Quote{M} \mid \AntiQuote{M} \mid \lift{M}
\end{equations}%
The grammar is largely the same as that of $\EffSub$. The key
difference is that effects are replaced by quotation
constructs. Lambda-abstractions are no longer annotated with
effects. More importantly, $\Quot$ includes quotation: we write
$\Quote{M}$ for the \emph{quotation} operation, where $M$ is a query
term; we write $\AntiQuote{M}$ for the \emph{antiquotation} operation,
which splices a quoted term $M$ into a query term; and we write
$\lift{M}$ for the operation that coerces a value of base type to a
quoted value.  (This is a limited form of \emph{cross-stage
  persistence}~\cite{taha00tcs}, which is otherwise unavailable
because quotation expressions must be closed.)  
%


Type environments $\Gamma,\Delta$ are as follows.
\[
\Gamma,\Delta ::= \cdot \mid \Gamma, x:A
\]

The typing rules are given in Figure~\ref{fig:quot-typing}. There are
two typing judgements: one for host terms, and the other for query
terms. The judgement $\Gamma \vdash M : A$ states that host term $M$
has type $A$ in type environment $\Gamma$. The judgement $\Gamma;
\Delta \vdash M : A$ states that query term $M$ has type $A$ in host
type environment $\Gamma$ and query type environment $\Delta$.

Most of the typing rules are standard and similar in both
judgements. The variable typing rule (\textsc{VarQ}) for the query
judgement does not allow the use of variables from the $\Gamma$
environment; hence variables from $\Gamma$ must be explicitly lifted
(rule \textsc{Lift}) or used within an antiquotation (rule
\textsc{Antiquote}) in order to be used within a quotation.

The interesting rules are those that involve quotation, namely
\textsc{Quote}, \textsc{Query}, \textsc{Antiquote}, and
\textsc{Lift}. Query terms can only be quoted in the host language
(\textsc{Quote}). In the host language, a closed quoted term of flat
bag type can be evaluated as a query (\textsc{Query}).
A quoted term can be spliced into a query term (\textsc{Antiquote}).
A host term of base type can be lifted to a query term
(\textsc{Lift}).  Note that variables of bag, record, or function type
in $\Gamma$ cannot be lifted using \textsc{Lift}.  As in $\EffSub$,
the $\Table~t$ construct is only available within a query; its use
within queries is enabled by rule \textsc{Table} and its use elsewhere
is forbidden because there is no similar rule for unquoted table
references.

\subsection{Operational Semantics for $\Quot$}

The operational semantics for $\Quot$ is standard, and is similar to
that given in Cheney et al.~\cite{cheney13icfp} for the \TLINQ
language.  
Values and evaluation contexts are given in
Figure~\ref{fig:quot-values}.  
The normalisation function $\normQuot{-}$ is the same as $\normEff{-}$
but applied to $\Quot$ instead of $\Eff$. Define $\normQuot{L} = N$ when $L
\rewriteto^* M$ and $M \adhocto^* N$, where $M$ and $N$ are in normal
form with respect to $\rewriteto$ and $\adhocto$ respectively.
%
The semantics is given in Figure~\ref{fig:quot-semantics}; most rules
are the same as in $\EffSub$, except those involving quotation and
querying. The rules are parameterised by the same $\delta$ and
$\Omega$ as the semantics of $\EffSub$.

\begin{proposition}
If\/ $\Gamma \vdash M : A$ and $M \ltoo{\mu} N$ then $\Gamma
\vdash N : A$.
If\/ $\Gamma \vdash M : A$ then either $M$ is a value or $M
\ltoo{\mu} N$ for some $N$ and $\mu$.
\end{proposition}

\begin{figure}
\begin{mysyntax}
\multicolumn{4}{@{}l@{}}{\cat{value}} \\
 & V,W &::=& c
    \mid \lambda x.M
    \mid \rec{f}{x}{M}
    \mid \record{\overline{\ell=V}}
    \mid \collection{\overline{V}}  \mid \Quote{Q} \\
\multicolumn{4}{@{}l@{}}{\cat{quotation value}} \\
 & Q &::=&  x \mid c \mid \op(\overline{Q})   \mid
    \lambda x.Q
    \mid Q~Q' \\
 && \mid & \If~Q~Q' \mid \record{\overline{\ell=Q}}
    \mid Q.\ell\\
 && \mid& \zero
    \mid \singleton{Q}
    \mid Q \plus Q' \mid \For\,(x \gets Q)\, Q' \mid \Table~t \\
\multicolumn{4}{@{}l@{}}{\cat{evaluation context}} \\
& \EC &::=& [~]
    \mid \op(\overline{V},\EC,\overline{M})
    \mid \lift{\EC}
    \mid \EC~ M \mid V~ \EC \\
 && \mid & \record{\overline{\ell=V},\ell'=\EC,\overline{\ell''=M}}
    \mid \EC.\ell
    \mid \singleton{\EC} \\
 && \mid & \EC \plus M
    \mid \singleton{V} \plus \EC
    \mid \For\,(x \gets \EC)\, N \\
 && \mid & \If~\EC~M
    \mid \Query~\EC
    \mid \Quote{\QC[\AntiQuote{\EC}]} \\
\multicolumn{4}{@{}l@{}}{\cat{quotation context}} \\
 & \QC  &::=& [~]
    \mid \op(\overline{Q},\QC,\overline{M})
    \mid \lambda x.\QC \\
 && \mid & \QC~ M
    \mid Q~\QC
    \mid \record{\overline{\ell=Q},\ell'=\QC,\overline{\ell''=M}} \\
 && \mid & \QC.\ell
    \mid \singleton{\QC}
    \mid \QC \plus M
    \mid Q \plus \QC \\
 && \mid & \For\,(x \gets \QC)\, N
    \mid \For\,(x \gets Q)\, \QC \\
 && \mid & \If~\QC~M
    \mid \If~Q~\QC
\end{mysyntax}%
\caption{Values and Evaluation Contexts for $\Quot$}
\label{fig:quot-values}


\begin{equations}
\op(\overline{V}) &\evalto& \opdelta(\op, \overline{V}) \\
(\lambda x.M)~ V &\evalto& M[x := V] \\
(\rec{f}{x}{M})~ V &\evalto& 
  M[f := \rec{f}{x}{M}, x := V] \\
\record{\overline{\ell=V}}.\ell_i &\evalto& V_i \\
\If~\True~M  &\evalto& M \\
\If~\False~M &\evalto& \zero \\
\zero \plus V &\evalto& V\\
(\singleton{V} \plus V') \plus V'' &\evalto& \singleton{V} \plus (V' \plus V'')\\
\fold\,\zero\,M\,N &\evalto& M \\
\fold\,((\singleton{V}) \plus W)\,M\,N &\evalto& N~V\,(\fold\,W\,M\,N) \\
\For\, (x \gets \zero)\, N &\evalto& \zero \\
\For\,(x \gets \singleton{V} \plus W)\, N &\evalto& N[x:=V]~\plus\\
&&\qquad (\For\, (x \gets W)\, N) \\
 \Query~\Quote{Q} &\ltoo{(\normQuot{Q},V)}& V\qquad\qquad\hfill ((\normQuot{Q},V) \in \Omega)  \\
 \lift{c} &\evaltopl& \Quote{c} \hfill (\text{lift}) \\
 \Quote{\QC[\AntiQuote{\Quote{Q}}]} &\evaltopl& \Quote{\QC[Q]} \hfill (\text{splice}) \\
\end{equations}%
\begin{mathpar}
\inferrule
{M \ltoo{\mu} N}{\EC[M] \ltoo{\mu} \EC[N]}
\end{mathpar}%

\caption{Operational Semantics for $\Quot$}
\label{fig:quot-semantics}
\end{figure}

\section{Translations}
\label{sec:translations}

\begin{figure}[tb]
\centering
  \includegraphics[scale=0.2]{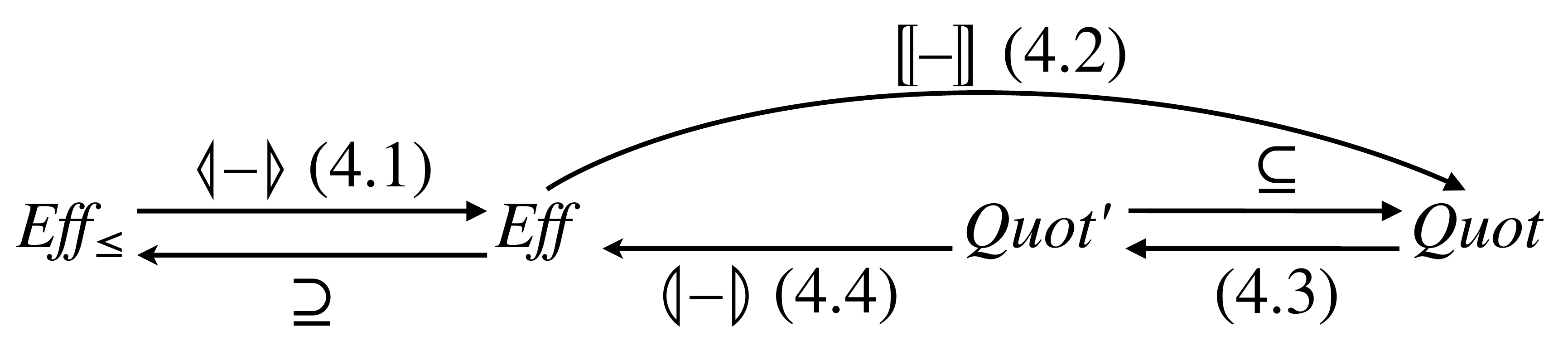}
  
  \caption{Summary of the Translations}
\label{fig:summary}
\end{figure}
In this section we present translations that show that $\EffSub$ and
$\Quot$ can simulate one another.  The translations are summarised in
Figure~\ref{fig:summary}.  We first (Section~\ref{sec:effsub-to-eff})
show how to compile away the $\any$-effect, translating arbitrary
$\EffSub$ programs to $\any$-free $\Eff$ programs.  Next
(Section~\ref{sec:eff-to-quot}) we show how to translate $\Eff$ programs
to $\Quot$ programs.  In the reverse direction, we first
(Section~\ref{sec:quot-to-quotprime}) give a straightforward translation hoisting computations out
of antiquotations in $\Quot$, resulting in a normal form $\Quot'$ in
which all antiquotations are of the form $\AntiQuote{x}$.  Finally
(Section~\ref{sec:quotprime-to-eff}) we translate $\Quot'$ programs to
$\Eff$ programs.

\subsection{From $\EffSub$ to $\Eff$}
\label{sec:effsub-to-eff}

As a first step, we can translate away all subeffecting through a
global \emph{doubling} translation that simulates each $\any$ function
as a pair of a $\pl$ function and a $\db$ function.  The type
translation $\dup{A}$ is shown in Figure~\ref{fig:doubling-type}.
Type environments are translated pointwise:
\[
\dup{x_1:A_1,\dots,x_n:A_n} = x_1:\dup{A_1},\dots,x_n:\dup{A_n}
\]
Terms are translated as shown in Figure~\ref{fig:doubling}, where
$\dupX{M}$ stands for the doubling translation of $M$ with respect to
target effect $X$.  The type and term translations are
structure-preserving except on $\any$-function types, abstractions,
and applications; these interesting cases are highlighted in grey
boxes. Technically, this translation is defined by induction on the
structure of typing derivations. However, the only cases where this
matters are those for applications of $\any$-functions.
%
%
To avoid notational clutter, we write the translation in a
syntax-directed style and only include type annotations in the cases for
application.

\begin{figure}
\small
\begin{equations}
  \dup{\Int}          &=& \Int \\
  \dup{\Bool}         &=& \Bool \\
  \dup{\String}    &=& \String \\
\dup{A \TO{X} B}  &=& \dup{A} \TO{X} \dup{B} \quad X \in \{\db,\pl\}\\
  \shaderow
  \dup{A \TO{\any} B} &=& \tuple{\dup{A} \TO{\pl} \dup{B}, \dup{A} \TO{\db} \dup{B}} \\
  \dup{\record{\overline{\ell:A}}}
  &=& \record{\overline{\ell:\dup{A}}} \\
  \dup{\tColl{A}}      &=& \tColl{\dup{A}} \\
\end{equations}
\caption{Doubling Translation: types}\label{fig:doubling-type}
\small
\begin{equations}
  \dupX{x}                  &=& x \\
  \dupX{c}                  &=& c \\
  \dupX{\op(\overline{M})}     &=& \op(\dupX{\overline{M}}) \\
  \dupX{\lambda^Y x^A.M}   &=& \lambda^Y x^{\dup{A}}.\dupY{M}  \quad
  Y \in \{\db,\pl\}\\
\shaderow
  \dupX{\lambda^\any x^A.M} &=& \langle
    \lambda^\pl x^{\dup{A}}.\duppl{M}, \lambda^\db
    x^{\dup{A}}.\dupdb{M} \rangle \\
  \dupX{M^{A \TO{X} B}~N}  &=& \dupX{M}~\dupX{N} \\
\shaderow
  \duppl{M^{A \TO{\any} B}~N} &=& \duppl{M}.1~\duppl{N} \\
\shaderow
  \dupdb{M^{A \TO{\any} B}~N} &=& \dupdb{M}.2~\dupdb{N} \\
  \dupX{\If\,L\,M}       &=& \If\,\dupX{L}~\dupX{M} \\
  \dupX{\record{\overline{\ell=M}}}
  &=& \record{\overline{\ell=\dupX{M}}} \\
  \dupX{M.\ell}             &=& \dupX{M}.\ell \\
  \dupX{\zero}            &=& \zero \\
  \dupX{\singleton{M}}      &=& \singleton{\dupX{M}} \\
  \dupX{M \plus N}          &=& \dupX{M} \plus \dupX{N} \\
  \dupX{\For\,(x^A \gets M)\,N}
  &=& \For\,(x^{\dup{A}} \gets \dupX{M})\,\dupX{N} \\
  \duppl{\rec{f^{A \TO{\pl} B}}{x^A}{M}}
  &=& \rec{f^{\dup{A \TO{\pl} B}}}{x^{\dup{A}}}{\duppl{M}} \\
  \duppl{\fold\,L\,M\,N} &=& \fold\,\duppl{L}\,\duppl{M}\,\duppl{N} \\
  \duppl{\Query\,M}       &=& \Query\,\dupdb{M} \\
  \dupdb{\Table~t}          &=& \Table~t \\
\end{equations}
\caption{Doubling Translation: terms}\label{fig:doubling}
\end{figure}

The term translation is parameterised by the concrete effect: $\pl$ or
$\db$. It is a straightforward structure-preserving traversal, except
on lambda-abstractions and applications of $\any$ functions. Each
$\any$-function is translated to a pair of a $\pl$-function and a
$\db$-function. To translate an application of an $\any$-function, the
target effect parameter is used to determine whether to use the first
or second element of the pair before applying the corresponding
function.

The main correctness properties are as follows:
\begin{theorem}[Type preservation]
  If $\Gamma \vdash M : A \eff E$ and $E \leq X$, then $\dup{\Gamma}
  \vdash \dupX{M} : \dup{A} \eff X$.
\end{theorem}
\begin{theorem}[$\any$-freedom]
  If $\Gamma \vdash M : A \eff E$ and $E \leq X$, then judgement
  $\dup{\Gamma} \vdash \dupX{M} : \dup{A} \eff X$ is derivable without
  $\any$.
\end{theorem}
The type preservation property is straightforward by structural
induction on derivations.  The $\any$-freedom property clearly holds
because the $\any$ effect appears nowhere on the right-hand-side of
the definitions of the $\dupX{-}$ functions.  An appropriate semantic
correctness property can also be shown, relating the operational
semantics of $\EffSub$ programs and their $\Eff$ translations.

\begin{theorem}
  [Semantics preservation]
Assume $\vdash M : A \eff \pl$.
  \begin{enumerate}
  \item $M$ is a value if and only if $\duppl{M}$ is a value. 
  \item If $M \ltoo{\mu} N$ then $\duppl{M} \lToo{\mu} \duppl{N}$.
  \item If $\duppl{M}$ is reducible then there exists $N$ such that
    $\duppl{M} \lToo{\mu} \duppl{N}$ and $M \ltoo{\mu} N$.
  \end{enumerate}
Moreover, if $\vdash M : A \eff \db$ then $\normEff{\dupdb{M}} = \dupdb{\normEff{M}}$.
\end{theorem}
\begin{corollary}
  For any $\vdash M: A \eff \pl$ in $\EffSub$ we have $M \trbisim \dup{M}_\pl$.
\end{corollary}

\subsection{From $\Eff$ to $\Quot$}
\label{sec:eff-to-quot}

We give an effect-directed translation from $\Eff$ to $\Quot$.  The
interpretations of types, terms and typing environments are
parameterised by the concrete effect: $\pl$ or $\db$.  The type
translation is written $\transX{A}$, where $X$ is $\pl$ or $\db$ and
$A$ is an $\Eff$-type, and is defined in
Figure~\ref{fig:splice-types}.  In the programming language, $\db$
function types are interpreted as quoted functions. In the database,
$\pl$ function types are interpreted as the unit type, which (as we
will show) suffices because $\pl$ functions can never be called
in the database.  

One additional complication is how to deal with occurrences of
variables bound within a $\pl$ context, that are also accessed by code
within a $\db$ context.  For example, consider the function
$\lambda^\pl x. \record{\lambda^\db y. \record{x,y},x}$: to translate
this, we need to be able to convert the value of $x$ to a quoted term,
but $\Quot$ only allows lifting at base type. We deal with this by
adding a special shadow variable $x_\db$ for each ordinary variable
$x$, so that the value of $x_\db$ is a quoted version of $x$.  In this
example, the translation is
\[\lambda x. \Let~x_\db = \Quote{\reify_A(x)}~\In~
\record{\Quote{\lambda y. \record{\AntiQuote{x_\db},y}},x}\;.\]
  The
special variables $x_\db$ and reification operation $\reify_A(-)$ are
explained in greater detail below.

To define the term translation, we need some auxiliary notation.  
The interpretations for terms and type environments are further
parameterised by an effect environment $\rho$, which tracks the
provenance of bound variables. This is necessary for interpreting a
value bound in a programming language context and used in a database context,
or vice-versa. The environment $\rho$ is a finite map from variable
names $x$ to pairs $A \eff X$ denoting the type $A$ and effect $X$ at
which $x$ was bound. We write $\varepsilon$ for the empty effect
environment and $\rho[x \mapsto A \eff X]$ for the extension of effect
environment $\rho$ with the mapping $x$ to $A \eff X$. 

We define a judgement $\Gamma \vdash \rho$ which states that effect
environment $\rho$ is compatible with type environment $\Gamma$.
\begin{mathpar}
  \inferrule[EmptyEnv] {~} {\cdot \vdash \varepsilon}

  \inferrule[ExtendEnv] {\Gamma \vdash \rho} {\Gamma,x:A \vdash \rho[x
    \mapsto A \eff X]}
\end{mathpar}%
The interpretation $\envAt{\rho}{X}$ of an effect environment $\rho$
at effect $X$ is defined as:
\begin{equations}
  \envAt{\varepsilon}{X}              &=& \cdot \\
  \envAt{\rho[x \mapsto A \eff \db]}{\db} &=& \envAt{\rho}{\db},x:\trans{A}_\db \\
  \envAt{\rho[x \mapsto A \eff \pl]}{\pl} &=&
  \envAt{\rho}{\pl},x:\trans{A}_\pl, x_\db:\trans{A}_\db \\
  \envAt{\rho[x \mapsto A \eff Y]}{X} &=& \envAt{\rho}{X} &\text{if }X \neq Y \\
\end{equations}%
which induces the interpretation of a type environment with respect to
a compatible effect environment. Notice that the translation
$\envAt{\rho}{\pl}$ introduces the shadow variables $x_\db$.  If $\Gamma
\vdash \rho$, then we define $\trans{\Gamma}_X^\rho$ as follows:
\[
\trans{\Gamma}_X^\rho = \envAt{\rho}{X}
\]


\begin{figure}
\small
\begin{equations}
  \transX{\Int}          &=& \Int \\
  \transX{\Bool}         &=& \Bool \\
  \transX{\String}    &=& \String \\
  \transX{A \TO{X} B}    &=& \transX{A} \to \transX{B} \\
\shaderow
  \transpl{A \TO{\db} B}  &=& \tQuote{\transdb{A \TO{\db} B}} \\
\shaderow
  \transdb{A \TO{\pl} B}  &=& \record{} \\
  \transX{\record{\overline{\ell:A}}} &=& \record{\overline{\ell:\transX{A}}} \\
  \transX{\tColl{A}}   &=& \tColl{\transX{A}} \\
\end{equations}%
\caption{Splicing Translation: types}
\label{fig:splice-types}
\begin{equations}
  \lookup{x}{\rho}{X} &=& x
  &\text{if }\rho(x) = A \eff X \\
  \lookup{x}{\rho}{\pl} &=& \error_{\transpl{A}}
  &\text{if }\rho(x) = A \eff \db \\
  \lookup{x}{\rho}{\db} &=& \AntiQuote{x_\db}
  &\text{if }\rho(x) = A \eff \pl \\
\end{equations}%
\begin{equations}
  \error_A &=& (\rec{f^{\record{} \to A}}{x^{\record{}}}{f~x})~\record{} \\
  \\
  \reify_O          ( M) &=& \AntiQuote{\lift{M}} \\
  \reify_{A \TO{\pl} B} (M) &=& \record{} \\
  \reify_{A \TO{\db} B} (M) &=& \AntiQuote{M} \\
  \reify_{\record{\overline{\ell:A}}} (M)
  &=& \record{\overline{\ell=\reify_A (M.\ell)}} \\
  \reify_{\tColl{A}} (M) &=& \\
  \multicolumn{3}{@{}l@{}}
  {\qquad\qquad\quad
   \AntiQuote{\fold~M~\Quote{\zero}~(\lambda x.\lambda y.\Quote{(\ret{\reify_A( x)}) \plus \AntiQuote{y}})}} \\
\end{equations}
\caption{Splicing Translation: auxiliary functions}\label{fig:splice-aux}
\end{figure}
\begin{figure}[tb]
\small
\begin{equations}
\shaderow
  \transXr{x}                &=& \lookup{x}{\rho}{X} \\
  \transXr{c}               &=& c \\
  \transXr{\op(\overline{M})}   &=& \op(\overline{\transXr{M}}) \\
\shaderow
  \transplr{\lambda^\pl x^A.M} &=& \lambda x^{\transpl{A}}.\\
\shaderow
&&\quad \Let~x_\db = \Quote{\reify_{A}(x)}~ \In\\
\shaderow
&&\quad\transpl{M}^{\rho[x \mapsto A \eff \pl]} \\
\shaderow
  \transplr{\lambda^\db x^A.M} &=& \Quote{\lambda x^{\transdb{A}}.\transdb{M}^{\rho[x \mapsto A \eff \db]}} \\
\shaderow
  \transdbr{\lambda^\pl x^A.M} &=& \record{} \\
\shaderow
  \transdbr{\lambda^\db x^A.M} &=& \lambda x^{\transdb{A}}.\transdb{M}^{\rho[x \mapsto A \eff \db]} \\
  \transXr{M~N} &=& \transXr{M}~\transXr{N} \\
  \transXr{\If\,L\,M}  &=& \If\,\transXr{L}~\transXr{M} \\
  \transXr{\record{\overline{\ell=M}}}
  &=& \record{\overline{\ell=\transXr{M}}} \\
  \transXr{M.\ell} &=& \transXr{M}.\ell \\
  \transXr{\zero} &=& \zero \\
  \transXr{\singleton{M}}      &=& \singleton{\transXr{M}} \\
  \transXr{M \plus N} &=& \transXr{M} \plus \transXr{N} \\
\shaderow   \transplr{\For\,(x^A \gets M)\,N} &=&
   \For\,(x^{\transpl{A}} \gets \transplr{M}) \\
\shaderow
&&(\Let~x_\db=\Quote{\reify_A(x)}~\In~\transpl{N}^{\rho[x \mapsto A \eff \pl]}) \\
\shaderow   \transdbr{\For\,(x^A \gets M)\,N} &=&
   \For\, (x^{\transdb{A}} \gets \transdbr{M}) \transdb{N}^{\rho[x \mapsto A \eff \db]} \\
\shaderow
  \transplr{\rec{f^{A \TO{\pl} B}}{x^A}{M}} &=& \rec{f^{\trans{A
        \TO{\pl} B}}}{x^{\transpl{A}}}{}\\
\shaderow
&&\quad\Let~f_\db =
  \Quote{\reify_{A\TO{\pl}B}(f)}~ \In\\
\shaderow
&&\quad\Let~x_\db =
  \Quote{\reify_{A}(x)} ~\In\\
\shaderow
&&\quad \transpl{M}^{\rho[f \mapsto (A \TO{\pl} B) \eff \pl, x \mapsto
  A \eff \pl]} \\
\shaderow
  \transdbr{\rec{f^{A \TO{\pl} B}}{x^A}{M}} &=& \record{} \\
  \transplr{\fold\,L\,M\,N} &=& \fold\,\transplr{L}\,\transplr{M}\,\transplr{N} \\
  \transplr{\Query~M} &=& \Query~\transdbr{M} \\
  \transdbr{\Table~t} &=& \Table~t \\
\end{equations}%
\caption{Splicing Translation: terms}
\label{fig:splice}
\end{figure}

The term translation is defined as $\transXr{M}$ in
Figure~\ref{fig:splice}, where $M$ is an $\Eff$ expression, $X$ is an
effect $\db$ or $\pl$ called the \emph{target} of the translation, and
$\rho$ is an effect environment mapping variables to their types and
effects (in $\Eff$).  The type translation is structure-preserving
except at function types, and similarly the term translation is
structure-preserving in most cases.  The interesting cases
(highlighted in grey) are those for function types,
lambda-abstractions, or variables where the translation's target
effect $X$ does not match the actual effect.  In addition, whenever a
variable $x$ is bound in the $\pl$ context, we bind an additional
special variable $x_\db$ whose value is a quoted version of $x$;
intuitively, we need this quoted value to translate any occurrences of
$x$ within a $\db$ context.  This affects all of the variable binding
cases of $\transplr{-}$ and is explained in more detail below.
Variables are coerced to effect $X$ using the $\lookup{x}{\rho}{X}$
operation described below.
Following the interpretation on types, $\lambda^\db$-abstractions are
coerced to a $\pl$ target by translating them to quoted
lambda-abstractions, and $\lambda^\pl$-abstractions are coerced to a
$\db$ target by translating them to unit values.

The coercion operation, written $\lookup{x}{\rho}{Y}$, is defined in
Figure~\ref{fig:splice-aux}. Given variable $x$ of type $\trans{A}_X$
(where $\rho(x) = A \eff X$), it yields a corresponding term in
$\Quot$ of type $\trans{A}_Y$.
If $X = Y$ then coercion leaves $x$ unchanged. If $X = \db$ and $Y =
\pl$, then an error term of type $\transpl{A}$ (implemented as a
diverging term) results. This is sound (but not strictly necessary)
because in a closed program a $\db$ variable can only be coerced to
$\pl$ inside the body of a $\pl$ function bound in a $\db$ context,
and such a function can never be applied. If $X = \pl$ and $Y = \db$,
then we splice in the value of $x_\db$, the special variable bound to
the \emph{reified value} of $x$ as a query term $\Quote{\reify_A(x)}$.

Reification is a type-directed operation that maps a term of type
$\transpl{A}$ to a corresponding term of type $\transdb{A}$. A term of
base type is reified by lifting and splicing. (Both are needed because
lifting coerces a value of base type to its quotation.) A $\pl$ function is reified as
unit. A $\db$ function is reified as an antiquotation. Records and
lists are reified by structural recursion on the type (which is why we
chose to include $\fold$ in the core calculi).



\begin{theorem}[Type preservation]
  \label{th:effquot-pres}
Assume $\Gamma \vdash \rho$.
  \begin{enumerate}
  \item\label{enum:effquot-pres-pl}
    If $\Gamma \vdash M : A \eff \pl$, then
    $\transplr{\Gamma}, \transdbr{\Gamma} \vdash \transplr{M} :
    \transpl{A}$.
  \item\label{enum:effquot-pres-db} 
    If $\Gamma \vdash M : A \eff \db$, then
    $\transplr{\Gamma}; \transdbr{\Gamma} \vdash \transdbr{M} :
    \transdb{A}$.
  \end{enumerate}
\end{theorem}

Notice that the translation of the database portion of the type
environment $\transdbr{\Gamma}$ appears in the programming language
judgement $\transplr{\Gamma}, \transdbr{\Gamma} \vdash \transplr{M} :
\transpl{A}$ in
Theorem~\ref{th:effquot-pres}(\ref{enum:effquot-pres-pl}). This is
sound as any such database variable in $\transplr{M}$ is interpreted
as $\error$. Database variables can never appear in a closed program.

\begin{theorem}[Semantics preservation]~
Assume $\vdash M : A \eff \pl$.  
\begin{enumerate}
\item  $M$ is a value if and only if
  $\transX{M}^\varepsilon$ is a value.  
\item If  $M
\ltoo{\mu} N$, then $\transpl{M}^\varepsilon \lToo{\mu} \transpl{N}^\varepsilon$.
\item If $\transpl{M}^\varepsilon$ is reducible
  then there exist $N$ and $\mu$ such that $\transpl{M}^\varepsilon \lToo{\mu}
  \transpl{N}^\varepsilon$ and $M \ltoo{\mu} N$.
\end{enumerate}
Moreover, if $\vdash M : A \eff \db$ then $\normQuot{\transdb{M}^\varepsilon} = \transdb{\normEff{M}}^\varepsilon$.
\end{theorem}

\begin{corollary}
  If $\vdash M : A \eff \pl$ in $\Eff$ then $M \trbisim
  \trans{M}^\varepsilon_\pl$.  
\end{corollary}

\begin{remark}
  The translation could be simplified by optimising away or inlining
  unnecessary $\Let$-bindings of $x_\db$ variables.  However, these
  simplifications complicate the correctness proof.
\end{remark}  

\subsection{From $\Quot$ to $\QuotVar$}
\label{sec:quot-to-quotprime}

In translating $\Quot$ to $\Eff$, the translation of an antiquote
presents a potential difficulty. It seems natural to translate an
antiquoted $\Quot$ term $\AntiQuote{M}$ into a corresponding $\Eff$
term with effect $\db$. The problem is that $M$ can perform arbitrary
computation including recursion. The solution is to hoist any spliced
computation out of the containing quotation. We can always soundly
hoist antiquoted computations out of quotations as they never depend
on the inner $\Delta$ environment. Thus as a preprocessing step we
replace all antiquoted terms with variables bound outside the scope of
the containing quotation. For convenience, we also perform similar
hoisting for lift and query expressions.  The target language of this
step, $\QuotVar$, is the restriction of $\Quot$ such that each
antiquotation, $\Query$, or $\lift{}$ may only be applied to a variable or value.

The hoisting translation uses the $\Let$ form (as usual) as syntactic
sugar for a lambda application:
\[
\Let\,x^A = M\,\In\,N \equiv (\lambda x^A.N)\,M
\] 
Hoisting is defined by repeatedly applying the following rules
\begin{equations}
\Quote{\QC[\AntiQuote{M}]} &\longrightarrow&
  \Let\,x=M\,\In\,\Quote{\QC[\AntiQuote{x}]} \\
\lift{M} &\longrightarrow & \Let\,x=M\,\In\,\lift{x}\\
\Query\,M &\longrightarrow&\Let\,x=M\,\In\,\Query\,x
\end{equations}%
where $M$ is not a variable or a value, and $x$ is a fresh
variable. 
Note that the structure of quotation contexts ensures that the terms
hoisted out of a quotation are still evaluated left-to-right with
respect to their original positions in the quotation.

We omit (routine but tedious) proofs of type-preservation and
semantics-preservation for this transformation.

\subsection{From $\QuotVar$ to $\Eff$}
\label{sec:quotprime-to-eff}

\begin{figure}[tb]\small
  \begin{equations}
    \untransX{\Int}    &=& \Int \\
    \untransX{\Bool}   &=& \Bool \\
\shaderow
    \untransX{A \to B} &=& \untransX{A} \TO{X} \untransX{B} \\
    \untransX{\record{\overline{\ell:A}}} &=& \record{\overline{\ell:\untransX{A}}} \\
    \untransX{\tColl{A}} &=& \tColl{\untransX{A}} \\
\shaderow
    \untransX{\tQuote{A}} &=& \record{} \TO{\db} \untransdb{A} \\
  \end{equations}%
  
  \caption{$\QuotVar$ to $\Eff$: types}
\label{fig:quot-to-eff-types}
  \begin{equations}
    \untransX{x}             &=& x \\
    \untransX{c}             &=& c \\
    \untransX{\op(\overline{M})} &=& \op(\overline{\untransX{M}}) \\
\shaderow
    \untransX{\lambda x^A.M} &=& \lambda^X x^{\untransX{A}}.\untransX{M} \\
    \untransX{M\,N}          &=& \untransX{M}\,\untransX{N} \\
    \untransX{\If\,L\,M}     &=& \If\,\untransX{L}~\untransX{M} \\
    \untransX{\record{\overline{\ell=M}}}
    &=& \record{\overline{\ell=\untransX{M}}} \\
    \untransX{M.\ell}        &=& \untransX{M}.\ell \\
    \untransX{\zero}         &=& \zero \\
    \untransX{\singleton{M}} &=& \singleton{\untransX{M}} \\
    \untransX{M \plus N}     &=& \untransX{M} \plus \untransX{N} \\
    \untransX{\For\,(x^A \gets M)\,N} &=&
    \For\,(x^{\untransX{A}} \gets \untransX{M})\,\untransX{N} \\
\shaderow    \untranspl{\lift{M}}
    &=& \lambda^\db x^{\record{}}. \untranspl{M} \\
\shaderow
\untranspl{\Quote{M}}
    &=& \lambda^\db x^{\record{}}. \untransdb{M} \\
\shaderow\untranspl{\Query\,M} &=& \Query\,(\untranspl{M}~\record{}) \\
    \untranspl{\rec{f^{A \to B}}{x^A}{M}}
    &=& \rec{f^{\untranspl{A \to B}}}{x^{\untranspl{A}}}{\untranspl{M}} \\
    \untranspl{\fold\,L\,M\,N}
    &=& \fold\,\untranspl{L}\,\untranspl{M}\,\untranspl{N} \\
    \untransdb{\Table~t}       &=& \Table~t \\
\shaderow   \untransdb{\AntiQuote{M}} &=& \untranspl{M}\,\record{} \\
  \end{equations}%
  
  \caption{$\QuotVar$ to $\Eff$: terms}
\label{fig:quot-to-eff}
\end{figure}
We now give a translation from $\QuotVar$ to $\Eff$.
 The type translation is shown in Figure~\ref{fig:quot-to-eff-types}.
It is structure-preserving except on closed quotation types, which are
translated to $\db$ thunks.

Type environments are translated pointwise:
\begin{equations}
\untransX{x_1:A_1,\dots,x_n:A_n} &=& x_1:\untransX{A_1},\dots,x_n:\untransX{A_n} \\
\end{equations}%

The term translation is shown in Figure~\ref{fig:quot-to-eff}.
It is structure-preserving, except on lambda-abstractions, quotation,
queries, antiquotation and lifting. Lambda-abstractions are annotated
with the appropriate effect. Quoted and lifted terms are translated to
$\db$ thunks $\lambda^\db x^\unit.\untransdb{M}$ or $\lambda^\db
x^\unit. \untranspl{M}$ respectively. For lifting, note that it does
not actually matter whether we use $\untranspl{-}$ or $\untransdb{-}$
on $M$, since in a $\QuotVar$ term, $M$ will always be either a
variable or constant of base type. Queries and antiquotations are
translated to force the $\db$ thunks by applying to unit.

\begin{remark}
The $\untransX{-}$ translation uses general rules for translating
  quotation, lifting, and antiquotation of arbitrary terms, but this
  is only correct for source expressions in $\QuotVar$.  The
  $\QuotVar$ terms resulting from the hoisting stage in the previous
  section will only have variables appearing as arguments to
  quotation, lifting, and antiquotation. However, this invariant is
  not preserved by evaluation, because variables may be replaced by values
  during evaluation. This is why $\QuotVar$ allows antiquotation,
  lifting, and query operations to be applied to values as well as
  variables.  
\end{remark}

\begin{theorem}[Type preservation]~
\begin{enumerate}
\item If $\Gamma \vdash M : A$, then $\untranspl{\Gamma} \vdash
  \untranspl{M} : \untranspl{A} \eff \pl$.

\item If $\Gamma; \Delta \vdash M : A$, then $\untranspl{\Gamma},
  \untransdb{\Delta} \vdash \untransdb{M} : \untransdb{A} \eff \db$.
\end{enumerate}
\end{theorem}

\begin{theorem}[Semantics preservation]~
Assume $\vdash M : A$.
\begin{enumerate}
\item If $M$ is a value then
  $\untranspl{M}$ is a value.
\item If $\Gamma \vdash M : A$ and $M \ltoo{\mu} N$, then $\untranspl{M}
  (\lToo{\mu} \cup \trbisim)  \untranspl{N}$.
\item If $\Gamma \vdash M : A$ and $\untranspl{M} $ is reducible then there exists $N$ and $\mu$
  such that $\untranspl{M} (\lToo{\mu} \cup \trbisim) \untranspl{N}$ and $M \ltoo{\mu} N$.
\end{enumerate}
Moreover, if $\cdot ; \cdot \vdash Q : A$ then $
\normEff{\untransdb{Q}} = \untransdb{\normQuot{Q}}$.
\end{theorem}

\begin{remark}
  Unlike the previous translations, these translations are not exact
  simulations of steps in $\Quot$ by steps in $\Eff$.  The reason is
  that the splicing rule in Figure~\ref{fig:quot-semantics}
  corresponds to $\beta$-value reduction under a $\lambda$-abstraction
  in $\Eff$. Fortunately, $\beta$-value equivalence is valid for
  $\Eff$ modulo $\trbisim$.
\end{remark}

\begin{corollary}
  If $\Gamma \vdash M : A$ in $\Quot$ then $M \trbisim \untranspl{M}$. 
\end{corollary}
\section{Application: Links query compilation}
\label{sec:implementation}

Holmes~\cite{Hol09} previously developed a compiler for plain Links.
Holmes' compiler translates Links programs (which include row polymorphism
and other features not present in plain OCaml) to OCaml programs that
manipulate tagged values. This typically improves the performance of
computationally-intensive Links programs by 1-2 orders of magnitude.
However, the compiler does not support Links' 
query or client-side Web programming features.

The translation from $\Eff$ to $\Quot$ shows how to
compile effect-based Links code to quotation-based 
code. Moreover, quotation-based queries are relatively easy to
translate to plain OCaml simply by translating quotation expressions
to an explicit run-time abstract syntax tree representation, and
extending the Links runtime library to include query normalisation.
Accordingly, we have adapted Holmes' compiler to support queries, by
first translating the Links intermediate representation (IR) code to
eliminate effect-polymorphic functions via doubling, then inserting
appropriate splicing annotations, and finally translating the
resulting IR code to OCaml code that explicitly manipulates quoted
terms.

The Links IR differs from $\EffSub$ in some important respects: in
particular, it employs row typing and effect polymorphism instead of
subeffecting. In the current implementation, we handle a subset of
Links and some polymorphic code is not handled. However, we believe
the basic idea of the doubling translation can be adapted to handle
polymorphism instead of subtyping. We view the current query compiler
as a proof of concept demonstrating practical implications of the
expressiveness results presented earlier; while it is not a mature
compiler for Links, our experience with it reported here will help
guide further development of such a compiler.

\begin{figure*}
\begin{center}
\begin{tabular}{ccc}
(a) map-$\any$&
(b) sumlist&
(c) quicksort\\
\includegraphics[scale=0.3]{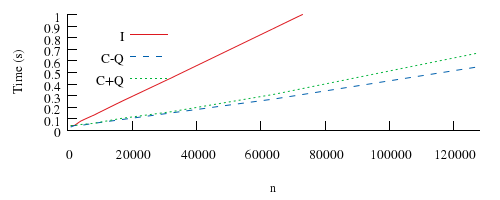}
&
\includegraphics[scale=0.3]{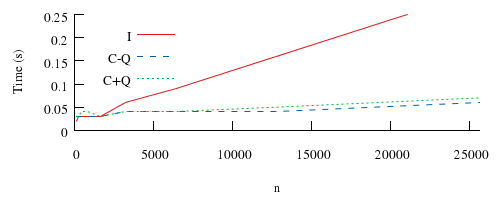}
&
\includegraphics[scale=0.3]{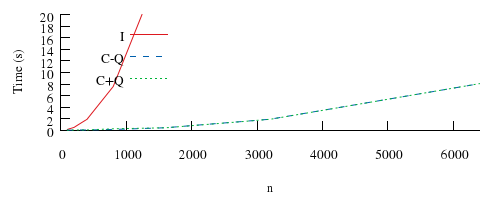}
\\
(d) static queries only&
(e) alternating static queries + sorting&
(f) alternating dynamic queries + sorting\\
\includegraphics[scale=0.3]{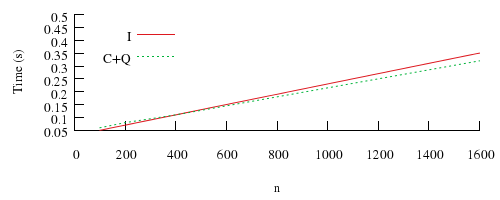}
&
\includegraphics[scale=0.3]{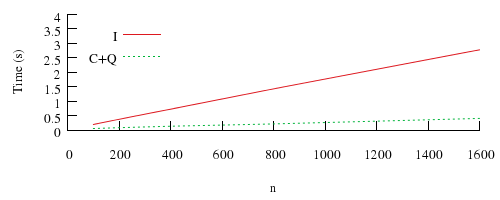}
&
\includegraphics[scale=0.3]{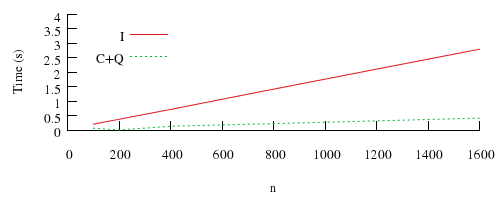}
\end{tabular}
\end{center}
\caption{Experimental Results}
\label{fig:experiments}
\end{figure*}

In the rest of this section we present some experiments showing the
strengths and weaknesses of the doubling and splicing translations,
which we hope will guide future work on compilation for
language-integrated query.  Figure~\ref{fig:experiments} compares the
interpreter (I), compiler without support for queries (C-Q) and
query-enabled compiler (C+Q) on several examples.  

We first consider programs that do not involve queries, and so can be
handled by all three techniques.  The map-$\any$ benchmark
(Figure~\ref{fig:experiments}(a)) measures the time needed to map a
simple $\any$-function (which calls some other $\any$-functions) over a
list.  The query-enabled compiler has a small, but measurable overhead
compared to the plain compiler.  The sumlist benchmark
(Figure~\ref{fig:experiments}(b)) measures the time to construct a
list of the first $n$ natural numbers, and compute its sum. Here, both
compilers provide similar results, though again C+Q is slightly
slower.  The quicksort benchmark (Figure~\ref{fig:experiments}(c))
sorts a decreasing $n$-element list, exercising the quadratic worst-case
behaviour of quicksort.  In this case, both compilers provide similar speedup.

We next consider examples that involve a mix of queries and ordinary
execution. These examples can only be run using the interpreter and
query-enabled compiler. The first benchmark
(Figure~\ref{fig:experiments}(d)) simply loops and generates $n$
queries, with no other computation. There is no measurable difference
between the two techniques. This is unsurprising since most of the
time is spent in communicating with the database, and compilation does
not affect this time, so speedup is bounded by Amdahl's Law.  Next, we
consider a variant (Figure~\ref{fig:experiments}(e)) where in each
iteration the program both issues a query and performs some
computation, namely a quicksort of a 10-element list. As we saw
before, the compiler yields a significant improvement on the quicksort
code. Finally, we consider a similar benchmark
(Figure~\ref{fig:experiments}(f)), where the query has a higher-order
parameter, which takes one value for odd-numbered iterations and
another for even-numbered iterations. The results of this benchmark
show no appreciable difference from Figure~\ref{fig:experiments}(e).

These results show that while the translation from $\Eff$ to $\Quot$
introduces some overhead to ordinary code compared to the basic
compiler, the query-enabled compiler still can realise significant
gains for code that mixes queries and ordinary execution.


\section{Related Work}
\label{sec:related}

There is a large and growing literature on approaches to
language-integrated query, as well as language-based techniques for
combining conventional execution with other execution models, such as
MapReduce, GPU, and multicore-based data-parallelism.  We discuss only
work closely related to this paper; other recent
papers~\cite{lindley12tldi,cheney13icfp} compare our approach to
language-integrated query with other work in
greater depth.

Wadler~\cite{Wadler92} advocated monads as a technique for structuring
programs, including list comprehensions and database queries.  This
approach was adopted in the nested relational calculus of Buneman et
al.~\cite{buneman+:comprehensions} and Wong~\cite{wong96jcss} gave
rewriting-based normalisation techniques for translating complex
nested relational queries over flat data to SQL queries, implemented
in the Kleisli system~\cite{Won00}.
Links~\cite{CLWY06,Cooper09,lindley12tldi} built on this work in
several ways, particularly in introducing the ability to compose
queries using nonrecursive lambda-abstractions and recursion in the
host language.

The LINQ approach adopted in C\#, F\# and other .NET languages also draws
upon monadic comprehensions and nested relational query
languages~\cite{meijer:sigmod,Meijer11}, but differs in its
implementation strategy: query syntax in C\# and F\# is desugared to
quoted abstract syntax trees which are manipulated and translated to
SQL by a library.  In F\#, it is possible to write dynamic LINQ
queries that fail at run time or generate unnecessarily large numbers
of SQL queries.  Our recent work~\cite{cheney13icfp} gave examples,
and showed how Links's normalisation algorithm can be adapted to F\#
to remedy this problem.  

Our formalisation of LINQ-style $\Quot$ draws upon a long line of work
on quotation and metaprogramming, starting with
MetaML~\cite{taha00tcs,Rhiger12}.  Our approach is closest to the
$\lambda^\Box$ calculus of Davies and Pfenning, which provides
homogeneous closed quotation (the host and quoted languages coincide);
for simplicity, we consider only one level of staging.  As discussed
elsewhere~\cite{cheney13icfp}, open quotation can be simulated in
$\Quot$ using lambda-abstraction, but better support for open
quotation and multiple stages, possibly following the approach of
Rhiger~\cite{Rhiger12}, may also be of interest.  Our approach also
has some similarities to Eckhardt et al.'s explicitly heterogeneous
approach~\cite{eckhardt07ngc}.  Reasoning about multi-stage programs
is a well-known hard problem. Choi et al.~\cite{Choi11} present
translations from staged to unstaged programs that employ similar
ideas to our translations, particularly hoisting $\pl$-code out of
$\db$-code in the $\untrans{-}$ translation.  Inoue and
Taha~\cite{inoue12esop} present techniques for reasoning about
call-by-value multi-stage programs.

Wadler and Thiemann demonstrate a close relationship between effect
type systems and monads~\cite{wadler-thiemann:marriage}. Our
translation from $\Eff$ to $\Quot$ has some similarities to that work.
However, the languages considered here are quite different from those
in Wadler and Thiemann's work; the latter employ reference types,
effects that are sets of regions, and monads indexed by sets of
regions, and there is nothing analogous to our doubling translation.

Felleisen~\cite{felleisen91scp} and Mitchell~\cite{mitchell93scp}
presented different notions of expressiveness of programming
languages, formulated in terms of different kinds of reductions
preserving termination behaviour or observational equivalence. We
adopt an ad hoc notion of equivalence based on preservation of query
behaviour, which is inspired to some extent by the notion of (weak)
bisimilarity familiar from concurrency theory~\cite{sangiorgi}.
However, in general nondeterministic labeled transition systems,
bisimilarity is strictly stronger than trace equivalence, so it is
possible that our translations do not preserve observable behaviour up
to bisimulation. We intend to investigate whether our translations are
(weak) bisimulations.

\section{Conclusion}
\label{sec:concl}

Combining database capabilities with general-purpose programming has
been of interest for nearly thirty years~\cite{copeland-maier:1984}.
Despite this long history, only within the last ten years have mature
techniques begun to appear in mainstream languages, with the chief
example being Microsoft's LINQ, based on explicitly manipulating query
code at run time using quotations.  Over the same period, techniques
developed in the Kleisli and Links languages have built on rigorous
foundations of query rewriting to show how to type-safely embed nested
relational queries in general-purpose languages, using type-and-effect
systems. 

In recent work~\cite{cheney13icfp}, we started to bring these threads
together, by showing that some techniques from Links, particularly
query normalisation, can be adapted to LINQ in F\# in order to improve
the expressiveness of the latter.  That work raised the question of
the relative expressiveness of the two approaches with respect to
dynamic query generation: Can Links express dynamic queries that LINQ
in principle cannot, or vice versa?

In this paper, we provided a partial answer to this question: we
proposed core languages $\EffSub$ and $\Quot$ similar to those used in
previous work on Links and LINQ respectively, and we gave
semantics-preserving translations in both directions.  This shows,
surprisingly in our view, that the two approaches are equivalent in
expressive power, at least relative to simple classes of queries: that
is, while Links programs or LINQ programs may seem more convenient in
different situations, in principle any program written using one
approach can also be written using the other.  In addition, we used
one direction of the translation to extend a Links compiler with
partial support for queries, demonstrating the effectiveness of
quotation for compiling Links.

A number of areas for future work remain, including extending our
translations to handle other query language features such as grouping,
aggregation and nested results; extending the compiler to handle full
Links including polymorphism; and completely eliminating the overhead
of doubling, which we believe should be possible using closure
conversion and storing each generated $\db$ function in a lookup table
indexed by the code pointer of the corresponding $\pl$ function.

\acks
This work is supported in part by a Google Research Award
(Lindley), EPSRC grants EP/J014591/1 (Lindley) and EP/K034413/1
(Lindley, Wadler), and a Royal Society University Research Fellowship
(Cheney).



\begin{thebibliography}{10}

\bibitem{ahv}
S.~Abiteboul, R.~Hull, and V.~Vianu.
\newblock {\em Foundations of Databases}.
\newblock Addison-Wesley, 1995.

\bibitem{buneman+:comprehensions}
P.~Buneman, L.~Libkin, D.~Suciu, V.~Tannen, and L.~Wong.
\newblock Comprehension syntax.
\newblock {\em SIGMOD Record}, 23, 1994.

\bibitem{BNTW95}
P.~Buneman, S.~Naqvi, V.~Tannen, and L.~Wong.
\newblock Principles of programming with complex objects and collection types.
\newblock {\em Theor. Comput. Sci.}, 149(1), 1995.

\bibitem{cheney13icfp}
J.~Cheney, S.~Lindley, and P.~Wadler.
\newblock A practical theory of language-integrated query.
\newblock In {\em ICFP}, 2013.

\bibitem{chlipala10pldi}
A.~J. Chlipala.
\newblock Ur: statically-typed metaprogramming with type-level record
  computation.
\newblock In {\em PLDI}, 2010.

\bibitem{Choi11}
W.~Choi, B.~Aktemur, K.~Yi, and M.~Tatsuta.
\newblock Static analysis of multi-staged programs via unstaging translation.
\newblock In {\em POPL}, pages 81--92. ACM, 2011.

\bibitem{Cooper09}
E.~Cooper.
\newblock The script-writer's dream: How to write great {SQL} in your own
  language, and be sure it will succeed.
\newblock In {\em DBPL}, 2009.

\bibitem{CLWY06}
E.~Cooper, S.~Lindley, P.~Wadler, and J.~Yallop.
\newblock Links: web programming without tiers.
\newblock In {\em FMCO}, 2007.

\bibitem{CooperW09}
E.~Cooper and P.~Wadler.
\newblock The {RPC} calculus.
\newblock In {\em PPDP}, 2009.

\bibitem{copeland-maier:1984}
G.~Copeland and D.~Maier.
\newblock Making {S}malltalk a database system.
\newblock {\em SIGMOD Rec.}, 14(2), 1984.

\bibitem{eckhardt07ngc}
J.~Eckhardt, R.~Kaiabachev, E.~Pasalic, K.~N. Swadi, and W.~Taha.
\newblock Implicitly heterogeneous multi-stage programming.
\newblock {\em New Generation Comput.}, 25(3):305--336, 2007.

\bibitem{felleisen91scp}
M.~Felleisen.
\newblock On the expressive power of programming languages.
\newblock {\em Sci. Comput. Programming}, 17:35--75, 1991.

\bibitem{dsh}
G.~Giorgidze, T.~Grust, T.~Schreiber, and J.~Weijers.
\newblock {Haskell} boards the {Ferry} - database-supported program execution
  for {Haskell}.
\newblock In {\em IFL}, number 6647 in LNCS, pages 1--18. Springer-Verlag,
  2010.

\bibitem{goldschmidt08icse}
T.~Goldschmidt, R.~Reussner, and J.~Winzen.
\newblock A case study evaluation of maintainability and performance of
  persistency techniques.
\newblock In {\em ICSE}, 2008.

\bibitem{grust13dbpl}
T.~Grust and A.~Ulrich.
\newblock First-class functions for first-order database engines.
\newblock In {\em DBPL}, 2013.
\newblock \texttt{http://arxiv.org/abs/1308.0158}.

\bibitem{henglein10hosc}
F.~Henglein and K.~F. Larsen.
\newblock Generic multiset programming with discrimination-based joins and
  symbolic cartesian products.
\newblock {\em Higher-Order and Symbolic Computation}, 23(3):337--370, 2010.

\bibitem{Hol09}
S.~Holmes.
\newblock Compiling {Links} server-side code.
\newblock Bachelor thesis, The University of Edinburgh, 2009.

\bibitem{inoue12esop}
J.~Inoue and W.~Taha.
\newblock Reasoning about multi-stage programs.
\newblock In {\em ESOP}, pages 357--376, 2012.

\bibitem{lindley12tldi}
S.~Lindley and J.~Cheney.
\newblock Row-based effect types for database integration.
\newblock In {\em Proceedings of the 8th ACM SIGPLAN workshop on Types in
  language design and implementation}, TLDI '12, 2012.

\bibitem{Meijer11}
E.~Meijer.
\newblock The world according to {LINQ}.
\newblock {\em Commun. ACM}, 54(10):45--51, Oct. 2011.

\bibitem{meijer:sigmod}
E.~Meijer, B.~Beckman, and G.~M. Bierman.
\newblock {LINQ}: reconciling object, relations and {XML} in the {.NET}
  framework.
\newblock In {\em SIGMOD}, 2006.

\bibitem{fsharp-query-expressions}
Microsoft.
\newblock Query expressions ({F\#} 3.0 documentation), 2013.
\newblock
  \texttt{http://msdn.microsoft.com/\-en-us/\-library/\-vstudio/\-hh225374.aspx},
  accessed March 18, 2013.

\bibitem{mitchell93scp}
J.~Mitchell.
\newblock On abstraction and the expressive power of programming languages.
\newblock {\em Sci. Comput. Programming}, 21:141--163, 1993.

\bibitem{petricek-dynamic-linq}
T.~Petricek.
\newblock Building {LINQ} queries at runtime in {C\#}, 2007.
\newblock \newline\texttt{http://tomasp.net/blog/dynamic-linq-queries.aspx}.

\bibitem{petricek-dynamic-flinq}
T.~Petricek.
\newblock Building {LINQ} queries at runtime in {F\#}, 2007.
\newblock \newline\texttt{http://tomasp.net/blog/dynamic-flinq.aspx}.

\bibitem{PetricekS14}
T.~Petricek and D.~Syme.
\newblock The {F\#} computation expression zoo.
\newblock In {\em PADL}, 2014.
\newblock To appear.

\bibitem{Rhiger12}
M.~Rhiger.
\newblock Staged computation with staged lexical scope.
\newblock In {\em ESOP}, number 7211 in LNCS, pages 559--578. Springer-Verlag,
  2012.

\bibitem{sangiorgi}
D.~Sangiorgi.
\newblock {\em Introduction to bisimulation and coinduction}.
\newblock Cambrudge University Press, 2012.

\bibitem{syme06}
D.~Syme.
\newblock Leveraging {.NET} meta-programming components from {F\#}: integrated
  queries and interoperable heterogeneous execution.
\newblock In {\em ML}, 2006.

\bibitem{fsharp3}
D.~Syme, A.~Granicz, and A.~Cisternino.
\newblock {\em Expert {F\#} 3.0}.
\newblock Apress, 2012.

\bibitem{taha00tcs}
W.~Taha and T.~Sheard.
\newblock {MetaML} and multi-stage programming with explicit annotations.
\newblock {\em Theor. Comput. Sci.}, 248(1-2):211--242, 2000.

\bibitem{talpin-jouvelot:effect-discipline}
J.-P. Talpin and P.~Jouvelot.
\newblock The type and effect discipline.
\newblock {\em Inf. and Comput.}, 111(2), 1994.

\bibitem{Wadler92}
P.~Wadler.
\newblock Comprehending monads.
\newblock {\em Math. Struct. in Comp. Sci.}, 2(4), 1992.

\bibitem{wadler-thiemann:marriage}
P.~Wadler and P.~Thiemann.
\newblock The marriage of effects and monads.
\newblock {\em Transactions on Computational Logic}, 4(1), 2003.

\bibitem{wassermann07tosem}
G.~Wassermann, C.~Gould, Z.~Su, and P.~Devanbu.
\newblock Static checking of dynamically generated queries in database
  applications.
\newblock {\em ACM Trans. Softw. Eng. Methodol.}, 16, September 2007.

\bibitem{wiedermann07popl}
B.~Wiedermann and W.~R. Cook.
\newblock Extracting queries by static analysis of transparent persistence.
\newblock In {\em POPL}, 2007.

\bibitem{wong96jcss}
L.~Wong.
\newblock Normal forms and conservative extension properties for query
  languages over collection types.
\newblock {\em J. Comput. Syst. Sci.}, 52(3), 1996.

\bibitem{Won00}
L.~Wong.
\newblock Kleisli, a functional query system.
\newblock {\em J. Funct. Program.}, 10(1), 2000.

\end{thebibliography}

\end{document}


